\documentclass[12pt]{article}
\usepackage{epsfig}
\newcommand\B{\rule[-1.2ex]{0pt}{0pt}}
      
      \def\di{\displaystyle}
      
      \def\bS{{\bf S}}
      \def\bl{{\bf l}}
      \def\bp{{\bf p}}
      \def\br{{\bf r}}
      \def\bs{{\bf s}}

      \def\L{{\cal L}}
      
      \def\P{{\cal P}}
      \def\R{{\cal R}}
      
      \def\Z{{\cal Z}}

\begin{document}

\begin{center}
{\large\bf 
SPIN SCISSORS MODE AND THE FINE STRUCTURE OF 
M1 STATES IN NUCLEI}\\
\vspace*{4mm}
{\large E.B. Balbutsev, I.V. Molodtsova}\\
%\vspace*{1mm}
{\it Joint Institute for Nuclear Research, 141980 Dubna, Moscow Region,
Russia}\\
\vspace*{1mm}
{\large P. Schuck}\\
%\vspace*{1mm}
{\it Institut de Physique Nucleaire, Orsay Cedex 91406, France}
\end{center}
%\vspace*{5mm}

\begin{abstract}

The coupled dynamics of low lying modes, including the scissors mode, and
various giant quadrupole resonances are studied with
the help of the Wigner Function Moments method generalized to
take into account spin degrees of freedom. Equations of motion for
collective variables are derived on the basis of Time Dependent
Hartree-Fock equations in the model of harmonic oscillator
with spin orbital mean field potential plus quadrupole-quadrupole residual
interaction. Introducing spin allows one to
consider new types of nuclear collective motion where the nucleons 
with spin 'up' oscillate against nucleons with spin 'down'. 

\end{abstract}

%      \vspace{5mm}

\section{Introduction}

The nuclear scissors mode was predicted \cite{Hilt,Hilt92,Suzuki,Lo} as the 
collective motion of two types of nucleons -- the protons undergo 
counter-rotational vibrations with respect to the neutrons. 
However, its collectivity turned out so small that
it was even discussed whether \cite{Zaw} this 
mode is really collective. Pure phenomenological models (such as, for 
example, two rotors model \cite{Lo2000}) could not help clarifying this
question. The results of RPA calculations \cite{Zaw} were
in qualitative agreement with the experiment,
an indication in favour of non collectivity. The final conclusion was
that the scissors mode is \cite{Heyd} "weakly collective, but strong 
on the single-particle scale". As a result \cite{Heyd}: "The weekly
collective scissors mode excitation has become an ideal test of models
-- especially microscopic models -- of nuclear vibrations. Most models
are usually calibrated to reproduce properties of strongly collective
excitations (e.g. of $J^{\pi}=2^+$ or $3^-$ states, giant resonances,
...). Weekly-collective phenomena, however, force the models to make
genuine predictions and the fact that the transitions in question are
strong on the single-particle scale makes it impossible to dismiss
failures as a mere detail, especially in the light of the overwhelming
experimental evidence for them in many nuclei \cite{Kneis,Richt}."

The Wigner Function
Moments (WFM) or phase space moments method turns out to be very 
useful in this
situation. On the one hand it is a pure microscopic method, because
it is based on the Time Dependent Hartree-Fock (TDHF) equation. On the
other hand the method works with average values (moments) of operators
which have a direct relation to the considered phenomenon. That 
makes it an ideal instrument to describe the basic characteristics 
(energies and excitation probabilities) of collective excitations such as,
in particular, the scissors mode. Our investigations have shown that 
already the minimal set of collective variables, i.e. phase space 
moments up to quadratic order,
is sufficient to reproduce the most important properties of the
scissors mode: its inevitable coexistence with the IsoVector Giant
Quadrupole Resonance (IVGQR) implying a deformation of the Fermi surface.
Also there is no doubt about its collectiveness,
because the principal variables generating this mode (angular momenta
of protons and neutrons) are collective being
the respective operators averaged over all protons and
neutrons.

Further developments of the Wigner Function Moments
method, namely, the switch from TDHF
to TDHFB equations, i.e. taking into account pair correlations, allowed
us to improve considerably the quantitative description of the 
scissors mode \cite{Malov,Urban}: for rare earth nuclei the energies are reproduced with
$\sim 10\%$ accuracy and B(M1) factors were reduced about two times. 
However, they remain about two times too high with respect to experiment.
The reason of the last discrepancy is probably hidden in the spin degrees 
of freedom, which were so far ignored by WFM method. One can not 
exclude, that due to spin dependent interactions some part of the 
force of M1 transitions is shifted to the energy region of 5-10 MeV,
where 1$^+$ resonance of the spin nature is observed. The 
generalization of the WFM method to take into account spin degrees of
freedom is the goal of this paper. In a first step, we include in the
consideration only the spin orbital interaction, as the most important
one among all possible spin dependent interactions because it enters 
into the mean field. This allows us to understand the structure
of necessary modifications of the method avoiding too cumbersome 
calculations. In this way it becomes clear already on the level of 
formulation of 
the equations of motion for new collective variables, that we are faced 
with a new type of collective motion, namely the spin scissors mode. 
It turns out that the experimentally 
observed group of peaks in the energy interval 2-4 MeV corresponds to
two different types of motion: "standard" scissors and this new kind 
of mode, i.e. spin scissors.

The paper is organized as follows.
In Sec. 2 the TDHF equations for 4-component density matrix are
formulated and their Wigner transform is found.
In Sec. 3 the spin structure of the density matrix and the model 
Hamiltonian are studied.
In Sec. 4 the collective variables are defined and the respective 
dynamical equations are derived.
In Sec. 5 the procedure of calculation of excitation probabilities is
recalled and the results of calculations of energies, B(M1) and B(E2)
factors are discussed.
Last remarks and the outlook are given in the conclusion section.

\section{TDHF equation with spin}

In this section we will consider the TDHF equation in coordinate 
space keeping all the spin indices. The TDHF equation
in operator formulation reads
\cite{Ring}
\begin{equation}
i\hbar\dot{\hat\rho}=[\hat h,\hat\rho]
\label{tHF}
\end{equation}
or in matrix form
\begin{eqnarray}
i\hbar<\br,s|\dot{\hat\rho}|\br'',s''> =
\hspace{12cm}
\nonumber\\
\hspace{10mm}
\sum_{s'}\int\!d^3r'(
<\br,s|\hat h|\br',s'><\br',s'|\hat\rho|\br'',s''> 
-<\br,s|\hat\rho|\br',s'><\br',s'|\hat h|\br'',s''>).
\label{HFmatr}
\end{eqnarray}
We do not specify the isospin indices in order 
to make the formulae more transparent. They
will be re-introduced at the end. 
The set of TDHF equations (\ref{HFmatr}) with specified spin indices reads
\begin{eqnarray}
&&i\hbar<\br|\dot{\hat\rho}|\br''>^{\uparrow\uparrow} =
\int\!d^3r'(
<\br|\hat h|\br'>^{\uparrow\uparrow}<\br'|\hat\rho|\br''>^{\uparrow\uparrow} 
-<\br|\hat\rho|\br'>^{\uparrow\uparrow}<\br'|\hat h|\br''>^{\uparrow\uparrow}
\nonumber\\
&&\hspace{40mm}
+<\br|\hat h|\br'>^{\uparrow\downarrow}<\br'|\hat\rho|\br''>^{\downarrow\uparrow} 
-<\br|\hat\rho|\br'>^{\uparrow\downarrow}<\br'|\hat h|\br''>^{\downarrow\uparrow}),
\nonumber\\
&&i\hbar<\br|\dot{\hat\rho}|\br''>^{\uparrow\downarrow} =
\int\!d^3r'(
<\br|\hat h|\br'>^{\uparrow\uparrow}<\br'|\hat\rho|\br''>^{\uparrow\downarrow} 
-<\br|\hat\rho|\br'>^{\uparrow\uparrow}<\br'|\hat h|\br''>^{\uparrow\downarrow}
\nonumber\\
&&\hspace{40mm}
+<\br|\hat h|\br'>^{\uparrow\downarrow}<\br'|\hat\rho|\br''>^{\downarrow\downarrow} 
-<\br|\hat\rho|\br'>^{\uparrow\downarrow}<\br'|\hat h|\br''>^{\downarrow\downarrow}),
\nonumber\\
&&i\hbar<\br|\dot{\hat\rho}|\br''>^{\downarrow\uparrow} =
\int\!d^3r'(
<\br|\hat h|\br'>^{\downarrow\uparrow}<\br'|\hat\rho|\br''>^{\uparrow\uparrow} 
-<\br|\hat\rho|\br'>^{\downarrow\uparrow}<\br'|\hat h|\br''>^{\uparrow\uparrow}
\nonumber\\
&&\hspace{40mm}
+<\br|\hat h|\br'>^{\downarrow\downarrow}<\br'|\hat\rho|\br''>^{\downarrow\uparrow} 
-<\br|\hat\rho|\br'>^{\downarrow\downarrow}<\br'|\hat h|\br''>^{\downarrow\uparrow}),
\nonumber\\
&&i\hbar<\br|\dot{\hat\rho}|\br''>^{\downarrow\downarrow} =
\int\!d^3r'(
<\br|\hat h|\br'>^{\downarrow\uparrow}<\br'|\hat\rho|\br''>^{\uparrow\downarrow} 
-<\br|\hat\rho|\br'>^{\downarrow\uparrow}<\br'|\hat h|\br''>^{\uparrow\downarrow}
\nonumber\\
&&\hspace{40mm}
+<\br|\hat h|\br'>^{\downarrow\downarrow}<\br'|\hat\rho|\br''>^{\downarrow\downarrow} 
-<\br|\hat\rho|\br'>^{\downarrow\downarrow}<\br'|\hat h|\br''>^{\downarrow\downarrow})
\label{HFsp}
\end{eqnarray}
with the conventional notation 
$$\uparrow \, \mbox{for}\quad s=\frac{1}{2} \quad \mbox{and}
\quad\downarrow \, \mbox{for}\quad s=-\frac{1}{2}.$$
These equations will be solved by the method of phase space (or Wigner
function) moments. To this end we will rewrite the expression (\ref{HFsp}) with the help of
Wigner transformation \cite{Ring}.

\subsection{ Wigner transformation}

The relevant mathematical details can be found in Appendix. 
To make the formulae more transparent we will not
write out the coordinate dependence $(\br,\bp)$ of the functions. The
Wigner transform of (\ref{HFsp}) can be written as
\begin{eqnarray}
      i\hbar\dot f^{\uparrow\uparrow} &=&i\hbar\{h^{\uparrow\uparrow},f^{\uparrow\uparrow}\}
+h^{\uparrow\downarrow}f^{\downarrow\uparrow}-f^{\uparrow\downarrow}h^{\downarrow\uparrow}
+\frac{i\hbar}{2}\{h^{\uparrow\downarrow},f^{\downarrow\uparrow}\}
-\frac{i\hbar}{2}\{f^{\uparrow\downarrow},h^{\downarrow\uparrow}\}
\nonumber\\
&&-\frac{\hbar^2}{8}\{\{h^{\uparrow\downarrow},f^{\downarrow\uparrow}\}\}
+\frac{\hbar^2}{8}\{\{f^{\uparrow\downarrow},h^{\downarrow\uparrow}\}\}+...,
\nonumber\\
      i\hbar\dot f^{\uparrow\downarrow} &=&
f^{\uparrow\downarrow}(h^{\uparrow\uparrow}-h^{\downarrow\downarrow})
+\frac{i\hbar}{2}\{(h^{\uparrow\uparrow}+h^{\downarrow\downarrow}),f^{\uparrow\downarrow}\}
-\frac{\hbar^2}{8}\{\{(h^{\uparrow\uparrow}-h^{\downarrow\downarrow}),f^{\uparrow\downarrow}\}\}
\nonumber\\&&
-h^{\uparrow\downarrow}(f^{\uparrow\uparrow}-f^{\downarrow\downarrow})
+\frac{i\hbar}{2}\{h^{\uparrow\downarrow},(f^{\uparrow\uparrow}+f^{\downarrow\downarrow})\}
+\frac{\hbar^2}{8}\{\{h^{\uparrow\downarrow},(f^{\uparrow\uparrow}-f^{\downarrow\downarrow})\}\}+....
\label{WHF}
\end{eqnarray}
where the functions $h$, $f$ are the Wigner transforms of $\hat h$, 
$\hat\rho$ respectively, $\{f,g\}$ is the Poisson
bracket of the functions $f(\br,\bp)$ and $g(\br,\bp)$ and
$\{\{f,g\}\}$ is their double Poisson bracket;
the dots stand for terms proportional to higher powers of $\hbar$.
Two more equations are obtained by the obvious change of arrows
$\uparrow \leftrightarrow \downarrow$.

It is useful to rewrite the above equations in terms of functions 
$f^+=f^{\uparrow\uparrow}+f^{\downarrow\downarrow}$,
$f^-=f^{\uparrow\uparrow}-f^{\downarrow\downarrow}$.
By analogy with isoscalar $f^{\rm n}+f^{\rm p}$ and isovector  
$f^{\rm n}-f^{\rm p}$ functions one can name the functions $f^+$ and 
$f^-$ as spin-scalar and spin-vector ones respectively.
We have:
\begin{eqnarray}
      i\hbar\dot f^{+} &=&\frac{i\hbar}{2}\{h^+,f^+\}+\frac{i\hbar}{2}\{h^-,f^-\}
+i\hbar\{h^{\uparrow\downarrow},f^d\}
+i\hbar\{h^{\downarrow\uparrow},f^u\}+...,
\nonumber\\
      i\hbar\dot f^{-} &=&
\frac{i\hbar}{2}\{h^+,f^-\}+\frac{i\hbar}{2}\{h^-,f^+\}
-2h^{\downarrow\uparrow}f^u+2h^{\uparrow\downarrow}f^d
\nonumber\\&&
+\frac{\hbar^2}{4}\{\{h^{\downarrow\uparrow},f^u\}\}
-\frac{\hbar^2}{4}\{\{h^{\uparrow\downarrow},f^d\}\}+...,
\nonumber\\
      i\hbar\dot f^u &=&
-h^{\uparrow\downarrow}f^-+h^-f^u
+\frac{i\hbar}{2}\{h^{\uparrow\downarrow},f^+\}
+\frac{i\hbar}{2}\{h^+,f^u\}
\nonumber\\&&
+\frac{\hbar^2}{8}\{\{h^{\uparrow\downarrow},f^-\}\}
-\frac{\hbar^2}{8}\{\{h^-,f^u\}\}+...,
\nonumber\\
      i\hbar \dot f^d &=&
h^{\downarrow\uparrow}f^- -h^-f^d
+\frac{i\hbar}{2}\{h^{\downarrow\uparrow},f^+\}
+\frac{i\hbar}{2}\{h^+,f^d\}
\nonumber\\&&
-\frac{\hbar^2}{8}\{\{h^{\downarrow\uparrow},f^-\}\}
+\frac{\hbar^2}{8}\{\{h^-,f^d\}\}+...,
\label{WHF+}
\end{eqnarray}
where $f^u=f^{\uparrow\downarrow},\,f^d=f^{\downarrow\uparrow},\,$ 
$h^{\pm}=h^{\uparrow\uparrow}\pm h^{\downarrow\downarrow}$.

\section{Model Hamiltonian}

The microscopic Hamiltonian of the model, harmonic oscillator with 
spin orbital interaction plus
separable quadrupole-quadrupole residual interaction is given by
\begin{eqnarray}
\label{Ham}
 H=\sum\limits_{i=1}^A[\frac{\hat\bp_i^2}{2m}+\frac{1}{2}m\omega^2\br_i^2
-\eta(\br_i)\hat \bl_i\hat \bS_i]
+\bar{\kappa}
\sum_{\mu=-2}^{2}(-1)^{\mu}
 \sum\limits_i^Z \sum\limits_j^N
q_{2-\mu}(\br_i)q_{2\mu}(\br_j)
\nonumber\\
+\frac{1}{2}\kappa
\sum_{\mu=-2}^{2}(-1)^{\mu}
\{\sum\limits_{i\neq j}^{Z}
 q_{2-\mu}(\br_i)q_{2\mu}(\br_j)
+\sum\limits_{i\neq j}^{N}
 q_{2-\mu}(\br_i)q_{2\mu}(\br_j)\},
\end{eqnarray}
where $N$ and $Z$ are the numbers of neutrons and protons, respectively.
The quadrupole operator $q_{2\mu}$ can be written as the tensor product:
$q_{2\mu}(\br)=\sqrt{16\pi/5}\,r^2Y_{2\mu}(\theta,\phi)=\sqrt6\{r\otimes r\}_{2\mu},$
where
$$
\{r\otimes r\}_{\lambda\mu}=\sum_{\sigma,\nu}
C_{1\sigma,1\nu}^{\lambda\mu}r_{\sigma}r_{\nu},$$
$r_{-1}, r_0, r_1$ are cyclic coordinates \cite{Var} and 
$C_{1\sigma,1\nu}^{\lambda\mu}$ is the Clebsch-Gordan
coefficient.
 The mean field potential for protons (or neutrons) is
\begin{equation}
\label{potenirr}
V^{\tau}=\frac{1}{2}m\,\omega^2r^2
+\sum_{\mu}(-1)^{\mu}Z_{2-\mu}^{\tau +}\{r\otimes r\}_{2\mu}
-\eta(\br)\hat \bl\hat \bS.
\end{equation}
 Here
$$
Z_{2\mu}^{\rm n+}=\chi R_{2\mu}^{\rm n+}
+\bar{\chi}R_{2\mu}^{\rm p+}\,,\quad
Z_{2\mu}^{\rm p+}=\chi R_{2\mu}^{\rm p+}
+\bar{\chi}R_{2\mu}^{\rm n+}\,,\quad
\chi=6\kappa,\quad\bar\chi=6\bar\kappa,$$
\begin{equation}
\label{Rlmu}
 R_{\lambda\mu}^{\tau+}(t)=
\int d\{\bp,\br\}
r_{\lambda\mu}^{2}f^{\tau+}(\br,\bp,t)
\end{equation}
 with
 $\int\! d\{\bp,\br\}\equiv
(2\pi\hbar)^{-3}\int\! d^3p\,\int\! d^3r$.

\subsection{Spin structure}

Let us clarify the spin structure of the density matrix and the
Hamiltonian.

\subsubsection{Density matrix}

As an example we consider the spherical case:
\begin{eqnarray}
\psi_{nljm}(\br,s)\equiv<nljm|\br,s>=\R_{nl}(r)
\sum_{\Lambda,\sigma}
C^{j m}_{l\Lambda,\frac{1}{2}\sigma}
 Y_{l\Lambda}(\theta,\phi)\chi_{\frac{1}{2}\sigma}(s),
\nonumber\\
\psi^*_{nljm}(\br,s)\equiv<\br,s|nljm>=\R_{nl}(r)\sum_{\Lambda,\sigma}
C^{j m}_{l\Lambda,\frac{1}{2}\sigma}
Y^*_{l\Lambda}(\theta,\phi)\chi^{\dagger}_{\frac{1}{2}\sigma}(s).
\label{phi}
\end{eqnarray}
Spin functions are defined as 
\begin{equation}
\chi_{\frac{1}{2}\frac{1}{2}}(s)={\delta_{s,\frac{1}{2}}\choose 0}
\equiv\chi_{\uparrow}(s),
\quad
\chi_{\frac{1}{2}-\frac{1}{2}}(s)={0\choose \delta_{s,-\frac{1}{2}}}
\equiv\chi_{\downarrow}(s),
\quad
\chi^{\dagger}_{\frac{1}{2}\frac{1}{2}}(s)=(\delta_{s,\frac{1}{2}}, 0).
\label{chi}
\end{equation}

The density matrix reads \cite{Ring}
\begin{eqnarray}
<\br,s|\hat \rho|\br',s'>&=&\sum_{nljm}v^2_{nlj}<\br,s|nljm><nljm|\br',s'>
\nonumber\\
&=&\sum_{nljm}v^2_{nlj}\R_{nl}(r)\R_{nl}(r')\sum_{\sigma_1,\sigma_2}
C^{j m}_{lm-\sigma_1,\frac{1}{2}\sigma_1}
C^{j m}_{lm-\sigma_2,\frac{1}{2}\sigma_2}
\nonumber\\
&&Y_{lm-\sigma_1}(\theta,\phi)Y^*_{lm-\sigma_2}(\theta',\phi')
\chi_{\frac{1}{2}\sigma_1}(s)\chi^{\dagger}_{\frac{1}{2}\sigma_2}(s'),
\label{rho}
\end{eqnarray}
where $v^2_{nlj}$ are the occupation numbers of level $nlj$.
According to the definition (\ref{chi}) the spin structure of the 
density matrix is 
\begin{equation}
<\br,s|\hat \rho|\br',s'>=\rho(\br s,\br' s'){\delta_{s\uparrow}\delta_{s'\uparrow}\quad 
\delta_{s\uparrow}\delta_{s'\downarrow}
\choose \delta_{s\downarrow}\delta_{s'\uparrow}\quad 
\delta_{s\downarrow}\delta_{s'\downarrow}},
\label{rhoS}
\end{equation}
where
\begin{eqnarray}
\rho(\br s,\br' s')=\sum_{nljm}v^2_{nlj}\phi_{nljm}(\br,s)\phi^*_{nljm}(\br',s')
\label{rhoss'}
\end{eqnarray}
and
\begin{eqnarray}
\phi_{nljm}(\br,s)
=\R_{nl}(r)C^{j m}_{lm-s,\frac{1}{2}s}Y_{lm-s}(\theta,\phi).
\label{phi1}
\end{eqnarray}

\subsubsection{l-s Hamiltonian}

Written in cyclic coordinates, the spin orbit part of the
Hamiltonian reads
$$\hat h_{ls}=-\eta(\br)\sum_{\mu=-1}^1(-)^{\mu}\hat l_{\mu}\hat S_{-\mu}
=-\eta(\br){\quad\hat l_0\frac{\hbar}{2}\quad\; \hat l_{-1}\frac{\hbar}{\sqrt2} \choose 
 -\hat l_{1}\frac{\hbar}{\sqrt2}\; -\hat l_0\frac{\hbar}{2}},
$$
where \cite{Var}
$$
\hat l_1=\hbar(r_0\nabla_1-r_1\nabla_0)=
\frac{\hbar}{\sqrt2}[x\nabla_z-z\nabla_x+i(y\nabla_z-z\nabla_y)]=
-\frac{1}{\sqrt2}(\hat l_x+i\hat l_y),$$
$$\hat l_0=\hbar(r_{-1}\nabla_1-r_1\nabla_{-1})=-i\hbar(x\nabla_y-y\nabla_x)=\hat l_z,$$
$$\hat l_{-1}=\hbar(r_{-1}\nabla_0-r_0\nabla_{-1})=
\frac{\hbar}{\sqrt2}[x\nabla_z-z\nabla_x-i(y\nabla_z-z\nabla_y)]=
\frac{1}{\sqrt2}(\hat l_x-i\hat l_y),
$$

\begin{equation}
\hat S_1=-\frac{\hbar}{\sqrt2}{0\quad 1\choose 0\quad 0},\quad
\hat S_0=\frac{\hbar}{2}{1\quad\, 0\choose 0\, -\!1},\quad
\hat S_{-1}=\frac{\hbar}{\sqrt2}{0\quad 0\choose 1\quad 0},
\label{S}
\end{equation}

The matrix elements of $\hat h_{ls}$ in the configuration space are:
\begin{eqnarray}
<\nu|\hat h_{ls}|\nu'>=\sum_{s,s'}\int\!d^3r<\nu|\br,s>h_{ls}<\br,s'|\nu'>
\nonumber\\
=-\sum_{s,s'}\int\!d^3r
\R_{nl}(r)
\sum_{\Lambda,\sigma}C^{jm}_{l\Lambda,\frac{1}{2}\sigma}
Y^*_{l\Lambda}(\theta,\phi)\chi^{\dagger}_{\frac{1}{2}\sigma}(s)
\nonumber\\
\eta(\br)\sum_{\mu}(-)^{\mu}\hat l_{\mu}(\br)\hat S_{-\mu}
\R_{n'l'}(r)
\sum_{\Lambda',\sigma'}C^{j'm'}_{l'\Lambda',\frac{1}{2}\sigma'}
Y_{l'\Lambda'}(\theta,\phi)\chi_{\frac{1}{2}\sigma'}(s'),
\label{Hnn'}
\end{eqnarray}

where $\nu\equiv n,l,j,m$. Taking into account, that
$$\chi^{\dagger}_{\uparrow}(s)\hat S_{1}\chi_{\uparrow}(s')=0,\quad
\chi^{\dagger}_{\uparrow}(s)\hat S_{-1}\chi_{\uparrow}(s')=0,\quad
\chi^{\dagger}_{\uparrow}(s)\hat S_{0}\chi_{\uparrow}(s')=\frac{\hbar}{2}
\delta_{s\frac{1}{2}}\delta_{s'\frac{1}{2}},$$

$$\chi^{\dagger}_{\downarrow}(s)\hat S_{1}\chi_{\downarrow}(s')=0,\quad
\chi^{\dagger}_{\downarrow}(s)\hat S_{-1}\chi_{\downarrow}(s')=0,\quad
\chi^{\dagger}_{\downarrow}(s)\hat S_{0}\chi_{\downarrow}(s')=-\frac{\hbar}{2}
\delta_{s-\frac{1}{2}}\delta_{s'-\frac{1}{2}},$$

$$\chi^{\dagger}_{\uparrow}(s)\hat S_{1}\chi_{\downarrow}(s')
=-\frac{\hbar}{\sqrt2}\delta_{s\frac{1}{2}}\delta_{s'-\frac{1}{2}},\quad
\chi^{\dagger}_{\uparrow}(s)\hat S_{-1}\chi_{\downarrow}(s')=0,\quad
\chi^{\dagger}_{\uparrow}(s)\hat S_{0}\chi_{\downarrow}(s')=0,$$

$$\chi^{\dagger}_{\downarrow}(s)\hat S_{1}\chi_{\uparrow}(s')=0,\quad
\chi^{\dagger}_{\downarrow}(s)\hat S_{-1}\chi_{\uparrow}(s')=
\frac{\hbar}{\sqrt2}\delta_{s-\frac{1}{2}}\delta_{s'\frac{1}{2}},\quad
\chi^{\dagger}_{\downarrow}(s)\hat S_{0}\chi_{\uparrow}(s')=0,$$

we find:
\begin{eqnarray}
<\nu|\hat h_{ls}|\nu'>=
-\sum_{s,s'}
\int\!d^3r
\sum_{\Lambda,\Lambda'}
\R_{nl}(r)Y^*_{l\Lambda}(\theta,\phi)\eta(\br)
\left[
C^{jm}_{l\Lambda,\frac{1}{2}\frac{1}{2}}
C^{j'm'}_{l'\Lambda',\frac{1}{2}\frac{1}{2}}
\frac{\hbar}{2}\hat l_{0}(\br)\delta_{s\frac{1}{2}}\delta_{s'\frac{1}{2}}
\hspace{2cm}
\right.
\nonumber\\
\left.
+C^{jm}_{l\Lambda,\frac{1}{2}\frac{1}{2}}
C^{j'm'}_{l'\Lambda',\frac{1}{2}-\frac{1}{2}}
\frac{\hbar}{\sqrt 2}\hat l_{-1}(\br)\delta_{s\frac{1}{2}}\delta_{s'-\frac{1}{2}}
-C^{jm}_{l\Lambda,\frac{1}{2}-\frac{1}{2}}
C^{j'm'}_{l'\Lambda',\frac{1}{2}\frac{1}{2}}
\frac{\hbar}{\sqrt 2}\hat l_{1}(\br)\delta_{s-\frac{1}{2}}\delta_{s'\frac{1}{2}}
\right.
\nonumber\\
\left.
-C^{jm}_{l\Lambda,\frac{1}{2}-\frac{1}{2}}
C^{j'm'}_{l'\Lambda',\frac{1}{2}-\frac{1}{2}}
\frac{\hbar}{2}\hat l_{0}(\br)\delta_{s-\frac{1}{2}}\delta_{s'-\frac{1}{2}}
\right]
\R_{n'l'}(r)Y_{l'\Lambda'}(\theta,\phi)
\nonumber\\
=-\sum_{s,s'}
\int\!d^3r
\sum_{\Lambda}
\R_{nl}(r)Y^*_{l\Lambda}(\theta,\phi)
C^{jm}_{l\Lambda,\frac{1}{2}s}\eta(\br)
\left[
\frac{\hbar}{2}\hat l_{0}(\br)\delta_{s\frac{1}{2}}\delta_{s'\frac{1}{2}}
+\frac{\hbar}{\sqrt 2}\hat l_{-1}(\br)\delta_{s\frac{1}{2}}\delta_{s'-\frac{1}{2}}
\right.
\nonumber\\
\left.
-\frac{\hbar}{\sqrt 2}\hat l_{1}(\br)\delta_{s-\frac{1}{2}}\delta_{s'\frac{1}{2}}
-\frac{\hbar}{2}\hat l_{0}(\br)\delta_{s-\frac{1}{2}}\delta_{s'-\frac{1}{2}}
\right]
\sum_{\Lambda'}
\R_{n'l'}(r)Y_{l'\Lambda'}(\theta,\phi)
C^{j'm'}_{l'\Lambda',\frac{1}{2}s'}
\nonumber\\
=-\sum_{s,s'}
\int\!d^3r
\phi^*_{nljm}(\br,s)\eta(\br)
\left[
\frac{\hbar}{2}\hat l_{0}(\br)\delta_{s\frac{1}{2}}\delta_{s'\frac{1}{2}}
+\frac{\hbar}{\sqrt 2}\hat l_{-1}(\br)\delta_{s\frac{1}{2}}\delta_{s'-\frac{1}{2}}
\hspace{3cm}
\right.
\nonumber\\
\left.
-\frac{\hbar}{\sqrt 2}\hat l_{1}(\br)\delta_{s-\frac{1}{2}}\delta_{s'\frac{1}{2}}
-\frac{\hbar}{2}\hat l_{0}(\br)\delta_{s-\frac{1}{2}}\delta_{s'-\frac{1}{2}}
\right]
\phi_{n'l'j'm'}(\br,s')
\nonumber\\
=<\nu|\hat h^{\uparrow\uparrow}|\nu'>+<\nu|\hat h^{\uparrow\downarrow}|\nu'>
+<\nu|\hat h^{\downarrow\uparrow}|\nu'>+<\nu|\hat h^{\downarrow\downarrow}|\nu'>,\hspace{2cm}
\label{Hnn'2}
\end{eqnarray}
where 
$$\hat h^{\uparrow\uparrow}(\br,s,s')=
-\frac{\hbar}{2}\eta(\br)\hat l_{0}(\br)\delta_{s\frac{1}{2}}\delta_{s'\frac{1}{2}},\quad
\hat h^{\uparrow\downarrow}(\br,s,s')=
-\frac{\hbar}{\sqrt 2}\eta(\br)\hat l_{-1}(\br)\delta_{s\frac{1}{2}}\delta_{s'-\frac{1}{2}},
$$
$$\hat h^{\downarrow\uparrow}(\br,s,s')=
\frac{\hbar}{\sqrt 2}\eta(\br)\hat l_{1}(\br)\delta_{s-\frac{1}{2}}\delta_{s'\frac{1}{2}},\quad
\hat h^{\downarrow\downarrow}(\br,s,s')=
\frac{\hbar}{2}\eta(\br)\hat l_{0}(\br)\delta_{s-\frac{1}{2}}\delta_{s'-\frac{1}{2}}.
$$
 
 Matrix elements of $\hat h_{ls}$ in coordinate space are derived
using the completeness of the set of functions $\phi_{nljm}(\br,s)$.
For example 
\begin{eqnarray}
<\br_1,s_1|\hat h^{\uparrow\uparrow}|\br_2,s_2>
=\sum_{\nu,\nu'}<\br_1,s_1|\nu><\nu|\hat h^{\uparrow\uparrow}|\nu'><\nu'|\br_2,s_2>
\nonumber\\
=-\frac{\hbar}{2}\sum_{nljm}\sum_{n'l'j'm'}\phi_{nljm}(\br_1,s_1)\sum_{s,s'}\int\!d^3r
\phi^*_{nljm}(\br,s)\eta(\br)\hat l_{0}(\br)\delta_{s\uparrow}\delta_{s'\uparrow}
\phi_{n'l'j'm'}(\br,s')\phi^*_{n'l'j'm'}(\br_2,s_2)
\nonumber\\
=-\frac{\hbar}{2}\sum_{s,s'}\int\!d^3r\delta(\br-\br_1)\delta_{s_1s}
\eta(\br)\hat l_{0}(\br)\delta_{s\uparrow}\delta_{s'\uparrow}\delta(\br-\br_2)\delta_{s_2s'}
=-\frac{\hbar}{2}\eta(\br_1)\hat l_{0}(\br_1)\delta(\br_1-\br_2)
\delta_{s_1\uparrow}\delta_{s_2\uparrow}
\nonumber\\
=\frac{\hbar}{2}\hat l_{0}(\br_2)[\eta(\br_2)\delta(\br_2-\br_1)]
\delta_{s_1\uparrow}\delta_{s_2\uparrow}
=h_{l_0}(\br_1,\br_2)\delta_{s_1\uparrow}\delta_{s_2\uparrow},
\label{Hrr'uu}
\end{eqnarray}
where
\begin{equation}
h_{l_0}(\br_1,\br_2)=-\frac{\hbar}{2}\eta(\br_1)\hat l_{0}(\br_1)\delta(\br_1-\br_2)
=\frac{\hbar}{2}\hat l_{0}(\br_2)[\eta(\br_2)\delta(\br_2-\br_1)].
\label{Hl0}
\end{equation}

Analogously we find
\begin{eqnarray}
<\br_1,s_1|\hat h^{\downarrow\downarrow}_{ss'}|\br_2,s_2>
=-h_{l_0}(\br_1,\br_2)\delta_{s_1\downarrow}\delta_{s_2\downarrow},
\nonumber\\
<\br_1,s_1|\hat h^{\uparrow\downarrow}_{ss'}|\br_2,s_2>
=h_{l_{-1}}(\br_1,\br_2)\delta_{s_1\uparrow}\delta_{s_2\downarrow},
\nonumber\\
<\br_1,s_1|\hat h^{\downarrow\uparrow}|\br_2,s_2>
=h_{l_{1}}(\br_1,\br_2)\delta_{s_1\downarrow}\delta_{s_2\uparrow}
\label{Hrr'dd}
\end{eqnarray}
with
\begin{equation}
h_{l_1}(\br_1,\br_2)=\frac{\hbar}{\sqrt 2}\eta(\br_1)\hat l_{1}(\br_1)\delta(\br_1-\br_2),\quad
h_{l_{-1}}(\br_1,\br_2)=-\frac{\hbar}{\sqrt 2}\eta(\br_1)\hat l_{-1}(\br_1)\delta(\br_1-\br_2).
\label{Hl1}
\end{equation}
As a result
\begin{eqnarray}
<\br_1,s_1|\hat h_{ls}|\br_2,s_2>
=h_{l_0}(\br_1,\br_2)[\delta_{s_1\uparrow}\delta_{s_2\uparrow}
-\delta_{s_1\downarrow}\delta_{s_2\downarrow}]
+h_{l_{-1}}(\br_1,\br_2)\delta_{s_1\uparrow}\delta_{s_2\downarrow}
+h_{l_{1}}(\br_1,\br_2)\delta_{s_1\downarrow}\delta_{s_2\uparrow}.
\label{Hrr'}
\end{eqnarray}
According to (\ref{WigH})--(\ref{WiH1}) its Wigner transformation reads:
\begin{eqnarray}
 h_{ls}(\br,\bp)
=-\frac{\hbar}{2}\eta(\br)\{l_{0}(\br,\bp)[\delta_{s_1\uparrow}\delta_{s_2\uparrow}
-\delta_{s_1\downarrow}\delta_{s_2\downarrow}]
+\sqrt2 l_{-1}(\br,\bp)\delta_{s_1\uparrow}\delta_{s_2\downarrow}
-\sqrt2 l_{1}(\br,\bp)\delta_{s_1\downarrow}\delta_{s_2\uparrow}\}.
\label{Hrp}
\end{eqnarray}

Taking into account 
equations (\ref{Hl0}) and (\ref{Hl1}) one can show that second 
and third equations of (\ref{HFsp}) are complex conjugated if the 
following relation is fulfilled:
$$<\br|\hat\rho|\br''>^{\uparrow\downarrow}=
<\br''|\hat\rho|\br>^{\downarrow\uparrow*},$$
i.e. if the density matrix is Hermitian as it must be.

\section{Equations of motion}

Integrating the set of equations (\ref{WHF+}) over the phase space 
with the weights 
\begin{equation}
W =\{r\otimes p\}_{\lambda\mu},\,\{r\otimes r\}_{\lambda\mu},\,
\{p\otimes p\}_{\lambda\mu} \mbox{ and } 1
\label{weightfunctions}
\end{equation}
one gets the dynamical equations for 
the following collective variables:
\begin{eqnarray}
L^{\pm}_{\lambda\mu}(t)=\int\! d(\bp,\br) \{r\otimes p\}_{\lambda\mu}
f^{\pm}(\br,\bp,t),\quad&&
\nonumber\\
L^{u}_{\lambda\mu}(t)=\int\! d(\bp,\br) \{r\otimes p\}_{\lambda\mu}
f^u(\br,\bp,t),\quad&&
\nonumber\\
L^{d}_{\lambda\mu}(t)=\int\! d(\bp,\br) \{r\otimes p\}_{\lambda\mu}
f^d(\br,\bp,t),\quad&&
\nonumber\\
R^{\pm}_{\lambda\mu}(t)=\int\! d(\bp,\br) \{r\otimes r\}_{\lambda\mu}
f^{\pm}(\br,\bp,t),\quad&&
\nonumber\\
R^{u}_{\lambda\mu}(t)=\int\! d(\bp,\br) \{r\otimes r\}_{\lambda\mu}
f^u(\br,\bp,t),\quad&&
\nonumber\\
R^{d}_{\lambda\mu}(t)=\int\! d(\bp,\br) \{r\otimes r\}_{\lambda\mu}
f^d(\br,\bp,t),\quad&&
\nonumber\\
P^{\pm}_{\lambda\mu}(t)=\int\! d(\bp,\br) \{p\otimes p\}_{\lambda\mu}
f^{\pm}(\br,\bp,t),\quad&&
\nonumber\\
P^{u}_{\lambda\mu}(t)=\int\! d(\bp,\br) \{p\otimes p\}_{\lambda\mu}
f^u(\br,\bp,t),\quad&&
\nonumber\\
P^{d}_{\lambda\mu}(t)=\int\! d(\bp,\br) \{p\otimes p\}_{\lambda\mu}
f^d(\br,\bp,t),\quad&&
\nonumber\\
F^{\pm}(t)=\int\! d(\bp,\br)
f^{\pm}(\br,\bp,t),\quad&&
\nonumber\\
F^{u}(t)=\int\! d(\bp,\br)
f^u(\br,\bp,t),\quad&&
\nonumber\\
F^{d}(t)=\int\! d(\bp,\br)
f^d(\br,\bp,t).\quad&&
\label{Varis}
\end{eqnarray}
We already named functions $f^{+}(\br,\bp,t)$ and $f^{-}(\br,\bp,t)$
as spin-scalar and spin-vector ones. So it will be natural to name the
respective collective variables $X^{+}_{\lambda\mu}(t)$ and 
$X^{-}_{\lambda\mu}(t)$ as spin-scalar and spin-vector variables.
The required expressions for $h^{\pm}$, $h^{\uparrow\downarrow}$
and $h^{\downarrow\uparrow}$ are
\begin{equation}
h^{+}=m\,\omega^2r^2
+2\sum_{\mu}(-1)^{\mu}Z_{2\mu}^{\tau+}\{r\otimes r\}_{2-\mu},\quad
h^-=-\hbar\eta l_0,\, h^{\uparrow\downarrow}=-\frac{\hbar}{\sqrt2}\eta l_{-1},
\, h^{\downarrow\uparrow}=\frac{\hbar}{\sqrt2}\eta l_{1}.
\label{hpm}
\end{equation}
The integration yields:
\begin{eqnarray}
\label{quadr}
     \dot L^{+}_{\lambda\mu}&=&
\frac{1}{m}P_{\lambda\mu}^{+}-
m\,\omega^2R^{+}_{\lambda \mu}
+2\sqrt5\sum_{j=0}^2\sqrt{2j+1}\{_{2\lambda 1}^{11j}\}
\{Z_2^{+}\otimes R_j^{+}\}_{\lambda \mu}
\nonumber\\
&&-i\hbar\frac{\eta}{2}[\mu L_{\lambda\mu}^- 
+\sqrt{(\lambda-\mu)(\lambda+\mu+1)}L^{u}_{\lambda\mu+1}+
\sqrt{(\lambda+\mu)(\lambda-\mu+1)}L^{d}_{\lambda\mu-1}],
\nonumber\\
     \dot L^{-}_{\lambda\mu}&=&
\frac{1}{m}P_{\lambda\mu}^{-}-
m\,\omega^2R^{-}_{\lambda \mu}
+2\sqrt5\sum_{j=0}^2\sqrt{2j+1}\{_{2\lambda 1}^{11j}\}
\{Z_2^{+}\otimes R_j^{-}\}_{\lambda \mu}
\nonumber\\&&
+i\eta\sqrt2\int\! d(\bp,\br)(rp)_{\lambda\mu}[l_1f^u+l_{-1}f^d]
\nonumber\\
&&-i\hbar\frac{\eta}{2}\mu L_{\lambda\mu}^+
-\frac{\hbar^2}{2}\eta\delta_{\lambda,1}
[\delta_{\mu,-1}F^u+\delta_{\mu,1}F^d],
\nonumber\\
     \dot L^{u}_{\lambda\mu+1}&=&
\frac{1}{m}P_{\lambda\mu+1}^{u}-
m\,\omega^2R^{u}_{\lambda \mu+1}
+2\sqrt5\sum_{j=0}^2\sqrt{2j+1}\{_{2\lambda 1}^{11j}\}
\{Z_2^{+}\otimes R_j^{u}\}_{\lambda \mu+1}
\nonumber\\&&
-i\frac{\eta}{\sqrt2}\int\! d(\bp,\br)(rp)_{\lambda\mu+1}[l_{-1}f^- -\sqrt2l_{0}f^u]
\nonumber\\
&&-i\hbar\frac{\eta}{4} \sqrt{(\lambda-\mu)(\lambda+\mu+1)}L_{\lambda\mu}^+
+\frac{\hbar^2}{2}\eta\delta_{\lambda,1}
[\delta_{\mu,0}F^- +\frac{1}{\sqrt2}\delta_{\mu,-1}F^u],
\nonumber\\
     \dot L^{d}_{\lambda\mu-1}&=&
\frac{1}{m}P_{\lambda\mu-1}^{d}-
m\,\omega^2R^{d}_{\lambda \mu-1}
+2\sqrt5\sum_{j=0}^2\sqrt{2j+1}\{_{2\lambda 1}^{11j}\}
\{Z_2^{+}\otimes R_j^{d}\}_{\lambda \mu-1}
\nonumber\\&&
-i\frac{\eta}{\sqrt2}\int\! d(\bp,\br)(rp)_{\lambda\mu-1}[l_1f^- +\sqrt2l_{0}f^d]
\nonumber\\
&&-i\hbar\frac{\eta}{4} \sqrt{(\lambda+\mu)(\lambda-\mu+1)}L_{\lambda\mu}^+
+\frac{\hbar^2}{4}\eta\delta_{\lambda,1}
[\delta_{\mu,0}F^- -\sqrt2\delta_{\mu,1}F^d],
\nonumber\\
     \dot F^{-}&=&
2\eta [L_{1-1}^d+L_{11}^u],
\nonumber\\
     \dot F^{u}&=&
-\eta [L_{1-1}^- -\sqrt2L_{10}^u],
\nonumber\\
     \dot F^{d}&=&
-\eta [L_{11}^- +\sqrt2L_{10}^d],
\nonumber\\
     \dot R^{+}_{\lambda\mu}&=&
\frac{2}{m}L^+_{\lambda\mu}
\nonumber\\
&&-i\hbar\frac{\eta}{2}[\mu R_{\lambda\mu}^- 
+\sqrt{(\lambda-\mu)(\lambda+\mu+1)}R^{u}_{\lambda\mu+1}+
\sqrt{(\lambda+\mu)(\lambda-\mu+1)}R^{d}_{\lambda\mu-1}],
\nonumber\\
     \dot R^{-}_{\lambda\mu}&=&
\frac{2}{m}L^-_{\lambda\mu}
-i\hbar\frac{\eta}{2}\mu R_{\lambda\mu}^+
+i\eta\sqrt2\int\! d(\bp,\br)(r^2)_{\lambda\mu}[l_1f^u+l_{-1}f^d],
\nonumber\\
     \dot R^{u}_{\lambda\mu+1}&=&
\frac{2}{m}L^u_{\lambda\mu+1}
%\nonumber\\&&
-i\frac{\eta}{\sqrt2}\int\! d(\bp,\br)(r^2)_{\lambda\mu+1}[l_{-1}f^- -\sqrt2l_{0}f^u]
\nonumber\\
&&-i\hbar\frac{\eta}{4} \sqrt{(\lambda-\mu)(\lambda+\mu+1)}R_{\lambda\mu}^+,
\nonumber\\
     \dot R^{d}_{\lambda\mu-1}&=&
\frac{2}{m}L^d_{\lambda\mu-1}
-i\frac{\eta}{\sqrt2}\int\! d(\bp,\br)(r^2)_{\lambda\mu-1}[l_1f^- +\sqrt2l_{0}f^d]
\nonumber\\
&&-i\hbar\frac{\eta}{4} \sqrt{(\lambda+\mu)(\lambda-\mu+1)}R_{\lambda\mu}^+,
\nonumber\\
     \dot P^{+}_{\lambda\mu}&=&
-2m\,\omega^2L^+_{\lambda \mu}
+4\sqrt5\sum_{j=0}^2\sqrt{2j+1}\{_{2\lambda 1}^{11j}\}
\{Z_2^+\otimes L^+_j\}_{\lambda \mu}
\nonumber\\
&&-i\hbar\frac{\eta}{2}[\mu P_{\lambda\mu}^- 
+\sqrt{(\lambda-\mu)(\lambda+\mu+1)}P^{u}_{\lambda\mu+1}+
\sqrt{(\lambda+\mu)(\lambda-\mu+1)}P^{d}_{\lambda\mu-1}],
\nonumber\\
     \dot P^{-}_{\lambda\mu}&=&
-2m\,\omega^2L^-_{\lambda \mu}
+4\sqrt5\sum_{j=0}^2\sqrt{2j+1}\{_{2\lambda 1}^{11j}\}
\{Z_2^{+}\otimes L^-_j\}_{\lambda \mu}
\nonumber\\&&
-i\hbar\frac{\eta}{2}\mu P_{\lambda\mu}^+
+i\eta\sqrt2\int\! d(\bp,\br)(p^2)_{\lambda\mu}[l_1f^u+l_{-1}f^d],
\nonumber\\
     \dot P^{u}_{\lambda\mu+1}&=&
-2m\,\omega^2L^u_{\lambda \mu+1}
+4\sqrt5\sum_{j=0}^2\sqrt{2j+1}\{_{2\lambda 1}^{11j}\}
\{Z_2^{+}\otimes L^u_j\}_{\lambda \mu+1}
\nonumber\\
&&-i\hbar\frac{\eta}{4} \sqrt{(\lambda-\mu)(\lambda+\mu+1)}P_{\lambda\mu}^+
-i\frac{\eta}{\sqrt2}\int\! d(\bp,\br)(p^2)_{\lambda\mu+1}[l_{-1}f^- -\sqrt2l_{0}f^u],
\nonumber\\
     \dot P^{d}_{\lambda\mu-1}&=&
-2m\,\omega^2L^d_{\lambda \mu-1}
+4\sqrt5\sum_{j=0}^2\sqrt{2j+1}\{_{2\lambda 1}^{11j}\}
\{Z_2^{+}\otimes L^d_j\}_{\lambda \mu-1}
\nonumber\\
&&-i\hbar\frac{\eta}{4} \sqrt{(\lambda+\mu)(\lambda-\mu+1)}P_{\lambda\mu}^+
-i\frac{\eta}{\sqrt2}\int\! d(\bp,\br)(p^2)_{\lambda\mu-1}[l_1f^- +\sqrt2l_{0}f^d],
\end{eqnarray}
 where $\{_{2\lambda 1}^{11j}\}$ is the Wigner
$6j$-symbol.
For the sake of simplicity the time dependence of tensors is not
written out. It is easy to see that this set of equations for moments
of zero and second orders is not closed due to the integral terms of
the type $
\int\! d(\bp,\br)(p^2)_{\lambda\mu-1}[l_1f^- +\sqrt2l_{0}f^d]
$ (see last equation). These terms generate moments of fourth 
order. They will be neglected according to the rules of the WFM method
\cite{Bal}.

We are interested in the scissors mode with quantum number
$K^{\pi}=1^+$. Therefore we only need the part of equations 
(\ref{quadr}) with $\mu=1$. These equations are nonlinear and will be
solved in small amplitude approximation.

\subsection{Linearized equations ($\mu=1$)} 

 Writing all variables as a sum of their
equilibrium value plus a small deviation
$$R_{\lambda\mu}(t)=R_{\lambda\mu}^{eq}+\R_{\lambda\mu}(t),\quad
P_{\lambda\mu}(t)=P_{\lambda\mu}^{eq}+\P_{\lambda\mu}(t),\quad
L_{\lambda\mu}(t)=L_{\lambda\mu}^{eq}+\L_{\lambda\mu}(t)$$
and neglecting quadratic deviations, one obtains the desired
equations. Naturally one needs to know the equilibrium values of all
variables.
Obvious equilibrium conditions for an axially symmetric nucleus are:
\begin{equation}
R^{+}_{2\pm1}(eq)=R^{+}_{2\pm2}(eq)=0,\quad R^{+}_{20}(eq)\neq0.
\label{equi1}
\end{equation}
It is obvious that all ground state properties of the system of spin
up nucleons are identical to that of the system of nucleons with spin
down. Therefore
\begin{equation}
R^{-}_{\lambda\mu}(eq)=P^{-}_{\lambda\mu}(eq)=L^{-}_{\lambda\mu}(eq)=0.
\label{equi2}
\end{equation}
We also will suppose 
\begin{equation}
L^{+}_{\lambda\mu}(eq)=L^{u}_{\lambda\mu}(eq)=L^{d}_{\lambda\mu}(eq)=0
\quad\mbox{ and }\quad
R^{u}_{\lambda\mu}(eq)=R^{d}_{\lambda\mu}(eq)=0.
\label{equi3}
\end{equation}
With the help of the above equilibrium relations one
arrives at the following set of equations:
\begin{eqnarray}
\label{mu1}
     \dot \L^{+}_{21}&=&
\frac{1}{m}\P_{21}^{+}-
m\,\omega^2\R^{+}_{21}
+\frac{2}{\sqrt3}R^{+}_{00}(eq)\Z^{+}_{21}
-\frac{1}{\sqrt6}[R^{+}_{20}(eq)\Z^{+}_{21}+Z^{+}_{20}(eq)\R^{+}_{21}]
\nonumber\\
&&-i\hbar\frac{\eta}{2}[\L_{21}^- 
+2\L^{u}_{22}+
2\sqrt{\frac{3}{2}}\L^{d}_{20}],
\nonumber\\
     \dot \L^{-}_{21}&=&
\frac{1}{m}\P_{21}^{-}-
m\,\omega^2\R^{-}_{21}
-\frac{1}{\sqrt6}Z^{+}_{20}(eq)\R^{-}_{21}
-i\hbar\frac{\eta}{2}\L_{21}^+ ,
\nonumber\\
     \dot \L^{u}_{22}&=&
\frac{1}{m}\P_{22}^{u}-
m\,\omega^2\R^{u}_{22}
-\sqrt{\frac{2}{3}}Z^{+}_{20}(eq)\R^{u}_{22}
-i\hbar\frac{\eta}{2}\L_{21}^+ ,
\nonumber\\
     \dot \L^{d}_{20}&=&
\frac{1}{m}\P_{20}^{d}-
m\,\omega^2\R^{d}_{20}
-\sqrt{\frac{2}{3}}Z^{+}_{20}(eq)\R^{d}_{20}
+\frac{2}{\sqrt3}Z^{+}_{20}(eq)\R^{d}_{00}
-i\hbar\frac{\eta}{2}\sqrt{\frac{3}{2}}\L_{21}^+ ,
\nonumber\\
     \dot \L^{+}_{11}&=&
\sqrt{\frac{3}{2}}[R^{+}_{20}(eq)\Z^{+}_{21}-Z^{+}_{20}(eq)\R^{+}_{21}]
-i\hbar\frac{\eta}{2}[\L_{11}^- 
+\sqrt2\L^{d}_{10}],
\nonumber\\
     \dot \L^{-}_{11}&=&
-\sqrt{\frac{3}{2}}Z^{+}_{20}(eq)\R^{-}_{21}
-i\hbar\frac{\eta}{2}\L_{11}^+ 
-\frac{\hbar^2}{2}\eta F^{d},
\nonumber\\
     \dot \L^{d}_{10}&=&
-i\hbar\frac{\eta}{2\sqrt2}\L_{11}^+ 
-\frac{\hbar^2}{4}\eta\sqrt2F^{d},
\nonumber\\
     \dot F^{d}&=&
-\eta[\L_{11}^- +\sqrt2\L^{d}_{10}],
\nonumber\\
     \dot \R^{+}_{21}&=&
\frac{2}{m}\L_{21}^{+}
-i\hbar\frac{\eta}{2}[\R_{21}^-
+2\R^{u}_{22}+
2\sqrt{\frac{3}{2}}\R^{d}_{20}],
\nonumber\\
     \dot \R^{-}_{21}&=&
\frac{2}{m}\L_{21}^{-}
-i\hbar\frac{\eta}{2}\R_{21}^+,
\nonumber\\
     \dot \R^{u}_{22}&=&
\frac{2}{m}\L_{22}^{u}
-i\hbar\frac{\eta}{2}\R_{21}^+,
\nonumber\\
     \dot \R^{d}_{20}&=&
\frac{2}{m}\L_{20}^{d}
-i\hbar\frac{\eta}{2}\sqrt{\frac{3}{2}}\R_{21}^+,
\nonumber\\
     \dot \P^{+}_{21}&=&
-2m\,\omega^2\L^{+}_{21}
+\sqrt6Z^{+}_{20}(eq)\L^{+}_{11}-\sqrt{\frac{2}{3}}Z^{+}_{20}(eq)\L^{+}_{21}
\nonumber\\
&&-i\hbar\frac{\eta}{2}[\P_{21}^- 
+2\P^{u}_{22}+2\sqrt{\frac{3}{2}}\P^{d}_{20}],
\nonumber\\
     \dot \P^{-}_{21}&=&
-2m\,\omega^2\L^{-}_{21}
+\sqrt6Z^{+}_{20}(eq)\L^{-}_{11}-\sqrt{\frac{2}{3}}Z^{+}_{20}(eq)\L^{-}_{21}
-i\hbar\frac{\eta}{2}\P_{21}^{+},
\nonumber\\
     \dot \P^{u}_{22}&=&
-2m\,\omega^2\L^{u}_{22}
+2\sqrt{\frac{2}{3}}Z^{+}_{20}(eq)\L^{u}_{22}
-i\hbar\frac{\eta}{2}\P_{21}^{+},
\nonumber\\
     \dot \P^{d}_{20}&=&
-2m\,\omega^2\L^{d}_{20}
-2\sqrt{\frac{2}{3}}Z^{+}_{20}(eq)\L^{d}_{20}
+2\sqrt{\frac{4}{3}}Z^{+}_{20}(eq)\L^{d}_{00}
-i\hbar\frac{\eta}{2}\sqrt{\frac{3}{2}}\P_{21}^{+},
\nonumber\\
     \dot \L^{d}_{00}&=&
\frac{1}{m}\P_{00}^{d}-m\,\omega^2\R^{d}_{00}
+\frac{2}{\sqrt3}Z^{+}_{20}(eq)\R^{d}_{20},
\nonumber\\
     \dot \R^{d}_{00}&=&
\frac{2}{m}\L_{00}^{d},
\nonumber\\
     \dot \P^{d}_{00}&=&
-2m\,\omega^2\L^{d}_{00}
+2\sqrt{\frac{4}{3}}Z^{+}_{20}(eq)\L^{d}_{20}.
\end{eqnarray}
Let us recall
that all variables and equilibrium quantities $R^{+}_{20}(eq)$
and $Z^{+}_{20}(eq)$ in (\ref{mu1}) have isospin indices 
$\tau=\rm n,\,\rm p$. All the difference 
between neutron and proton systems lies in the mean field 
quantity $Z^{\tau+}_{20}(eq)$, which
is different for neutrons and protons (see (\ref{Rlmu})).

\subsection{Isovector, isoscalar}

  It is convenient to rewrite the equations (\ref{mu1}) in terms
of isovector and isoscalar variables
$$R_{\lambda\mu}=R_{\lambda\mu}^{\rm n}+R_{\lambda\mu}^{\rm p},\quad
P_{\lambda\mu}=P_{\lambda\mu}^{\rm n}+P_{\lambda\mu}^{\rm p},\quad
L_{\lambda\mu}=L_{\lambda\mu}^{\rm n}+L_{\lambda\mu}^{\rm p},$$
$$\bar R_{\lambda\mu}=R_{\lambda\mu}^{\rm n}-R_{\lambda\mu}^{\rm p}
,\quad
\bar P_{\lambda\mu}=P_{\lambda\mu}^{\rm n}-P_{\lambda\mu}^{\rm p}
,\quad
\bar L_{\lambda\mu}=L_{\lambda\mu}^{\rm n}-L_{\lambda\mu}^{\rm p}.$$
It is natural to define also isovector and isoscalar strength constants
$\chi_1=\frac{1}{2}(\chi-\bar\chi)$ and
$\chi_0=\frac{1}{2}(\chi+\bar\chi)$ connected by the relation
$\chi_1=\alpha\chi_0$ \cite{BaSc}.
Then the equations for the neutron and proton systems are transformed
into isovector and isoscalar ones.
The equations for the isovector system are given by
\begin{eqnarray}
\label{iv}
     \dot {\bar\L}^{+}_{21}&=&
\frac{1}{m}\bar\P_{21}^{+}-
\left[m\,\omega^2
-\frac{2}{\sqrt3}\alpha\chi_0R^{+}_{00}(eq)
+\frac{1}{\sqrt6}(1+\alpha)\chi_0 R^{+}_{20}(eq)\right]\bar\R^{+}_{21}
\nonumber\\
&&-i\hbar\frac{\eta}{2}[\bar\L_{21}^-
+2\bar\L^{u}_{22}+
2\sqrt{\frac{3}{2}}\bar\L^{d}_{20}],
\nonumber\\
     \dot {\bar\L}^{-}_{21}&=&
\frac{1}{m}\bar\P_{21}^{-}-
[m\,\omega^2+\frac{1}{\sqrt6}\chi_0 R_{20}(eq)]\bar\R^{-}_{21}
-i\hbar\frac{\eta}{2}\bar\L_{21}^+ ,
\nonumber\\
     \dot {\bar\L}^{u}_{22}&=&
\frac{1}{m}\bar\P_{22}^{u}-
[m\,\omega^2
-\sqrt{\frac{2}{3}}\chi_0R_{20}(eq)]\bar\R^{u}_{22}
-i\hbar\frac{\eta}{2}\bar\L_{21}^+ ,
\nonumber\\
     \dot {\bar\L}^{d}_{20}&=&
\frac{1}{m}\bar\P_{20}^{d}-
[m\,\omega^2
+\sqrt{\frac{2}{3}}\chi_0 R_{20}(eq)]\bar\R^{d}_{20}
+\frac{2}{\sqrt3}\chi_0 R_{20}(eq)\bar\R^{d}_{00}
-i\hbar\frac{\eta}{2}\sqrt{\frac{3}{2}}\bar\L_{21}^+ ,
\nonumber\\
     \dot {\bar\L}^{+}_{11}&=&
-\sqrt{\frac{3}{2}}(1-\alpha)\chi_0 R_{20}(eq)\bar\R^{+}_{21}
-i\hbar\frac{\eta}{2}[\bar\L_{11}^- 
+\sqrt2\bar\L^{d}_{10}],
\nonumber\\
     \dot {\bar\L}^{-}_{11}&=&
-\sqrt{\frac{3}{2}}\chi_0 R_{20}(eq)\bar\R^{-}_{21}
-\hbar\frac{\eta}{2}[i\bar\L_{11}^+
+\hbar\bar F^{d}],
\nonumber\\
     \dot {\bar\L}^{d}_{10}&=&
-\hbar\frac{\eta}{2\sqrt2}[i\bar\L_{11}^+
+\hbar\bar F^{d}],
\nonumber\\
     \dot {\bar F}^{d}&=&
-\eta[\bar\L_{11}^- +\sqrt2\bar\L^{d}_{10}],
\nonumber\\
     \dot {\bar\R}^{+}_{21}&=&
\frac{2}{m}\bar\L_{21}^{+}
-i\hbar\frac{\eta}{2}[\bar\R_{21}^-
+2\bar\R^{u}_{22}+
2\sqrt{\frac{3}{2}}\bar\R^{d}_{20}],
\nonumber\\
     \dot {\bar\R}^{-}_{21}&=&
\frac{2}{m}\bar\L_{21}^{-}
-i\hbar\frac{\eta}{2}\bar\R_{21}^+,
\nonumber\\
     \dot {\bar\R}^{u}_{22}&=&
\frac{2}{m}\bar\L_{22}^{u}
-i\hbar\frac{\eta}{2}\bar\R_{21}^+,
\nonumber\\
     \dot {\bar\R}^{d}_{20}&=&
\frac{2}{m}\bar\L_{20}^{d}
-i\hbar\frac{\eta}{2}\sqrt{\frac{3}{2}}\bar\R_{21}^+,
\nonumber\\
     \dot {\bar\P}^{+}_{21}&=&
-[2m\,\omega^2+\sqrt{\frac{2}{3}}\chi_0 R_{20}(eq)]\bar\L^{+}_{21}
+\sqrt6\chi_0 R_{20}(eq)\bar\L^{+}_{11}
%\nonumber\\&&
-i\hbar\frac{\eta}{2}[\bar\P_{21}^- 
+2\bar\P^{u}_{22}+2\sqrt{\frac{3}{2}}\bar\P^{d}_{20}],
\nonumber\\
     \dot {\bar\P}^{-}_{21}&=&
-[2m\,\omega^2+\sqrt{\frac{2}{3}}\chi_0 R_{20}(eq)]\bar\L^{-}_{21}
+\sqrt6\chi_0 R_{20}(eq)\bar\L^{-}_{11}
-i\hbar\frac{\eta}{2}\bar\P_{21}^{+},
\nonumber\\
     \dot {\bar\P}^{u}_{22}&=&
-[2m\,\omega^2-2\sqrt{\frac{2}{3}}\chi_0 R_{20}(eq)]\bar\L^{u}_{22}
-i\hbar\frac{\eta}{2}\bar\P_{21}^{+},
\nonumber\\
     \dot {\bar\P}^{d}_{20}&=&
-[2m\,\omega^2+2\sqrt{\frac{2}{3}}\chi_0 R_{20}(eq)]\bar\L^{d}_{20}
+\frac{4}{\sqrt3}\chi_0 R_{20}(eq)\bar\L^{d}_{00}
-i\hbar\frac{\eta}{2}\sqrt{\frac{3}{2}}\bar\P_{21}^{+},
\nonumber\\
     \dot {\bar\L}^{d}_{00}&=&
\frac{1}{m}\bar\P_{00}^{d}-m\,\omega^2\bar\R^{d}_{00}
+\frac{2}{\sqrt3}\chi_0 R_{20}(eq)\bar\R^{d}_{20},
\nonumber\\
     \dot {\bar\R}^{d}_{00}&=&
\frac{2}{m}\bar\L_{00}^{d},
\nonumber\\
     \dot {\bar\P}^{d}_{00}&=&
-2m\,\omega^2\bar\L^{d}_{00}
+\frac{4}{\sqrt3}\chi_0 R_{20}(eq)\bar\L^{d}_{20}.
\end{eqnarray}
The isoscalar set of equations is easily obtained from (\ref{iv}) by 
taking $\alpha=1$.

\subsection{Angular momentum conservation}

The set of equations (\ref{iv}) contains three integrals of motion
(see Appendix B). 
The first one is (in the isoscalar case) the conservation of total 
angular momentum
$<\hat J_1>=<\hat l_1>+<\hat S_1>$. By definition
\begin{eqnarray}
\label{l1}
<\hat l_1>=Tr(\hat l_1 \hat\rho)=
\sum_s\int d^3r\int d^3r'<\br|\hat l_1|\br'><\br',s|\hat\rho|\br,s>
\nonumber\\
=\sum_s\int d^3r\int d^3r'\hat l_1(\br)\delta(\br-\br')<\br',s|\hat\rho|\br,s>
=\int d^3r\hat l_1(\br)[<\br|\hat\rho|\br>^{\uparrow\uparrow}+
<\br|\hat\rho|\br>^{\downarrow\downarrow}]
\nonumber\\
=\int d(\bp\br)l_1(\br,\bp)f^+(\br,\bp)
=-i\sqrt2\int d(\bp\br)\{r\otimes p\}_{11}f^+(\br,\bp)=-i\sqrt2L^+_{11}.
\end{eqnarray}
The average value of the spin operator $\hat S_1$ reads:
\begin{eqnarray}
\label{S1}
<\hat S_1>=Tr(\hat S_1 \hat\rho)=
\sum_{s,s'}\int d^3r<s|\hat S_1|s'><\br,s'|\hat\rho|\br,s>
\nonumber\\
=\sum_{s,s'}<s|\hat S_1|s'>\int d(\bp\br)f^{s's}(\br,\bp)
=-\frac{\hbar}{\sqrt2}\sum_{s,s'}\delta_{s\uparrow}\delta_{s'\downarrow}F^{s's}
=-\frac{\hbar}{\sqrt2}F^{\downarrow\uparrow}
\equiv -\frac{\hbar}{\sqrt2}F^{d}.
\end{eqnarray}
As a result $<\hat J_1>=-\frac{1}{\sqrt2}(2iL^+_{11}+\hbar F^{d})$. 
It is easy to see that such combination of the respective equations in 
(\ref{iv}) is equal to zero in the isoscalar case $(\alpha=1)$,
i.e. the total angular momentum is conserved.

\section{Energies and excitation probabilities}

Imposing the time evolution via $\di{e^{i\Omega t}}$ for all variables
one transforms (\ref{iv}) into a set of algebraic equations. 
Eigenfrequencies are found as the zeros of its secular equation. Excitation
probabilities are calculated with the help of the theory of linear 
response of the system to a weak external field
\begin{equation}
\label{extf}
\hat O(t)=\hat O\,exp(-i\Omega t)+\hat O^{\dagger}\,exp(i\Omega t).
\end{equation}
The detailed explanation can be found in \cite{BaSc}. Here we only 
will recall the main points.
The matrix elements of the operator $\hat O$ obey the relationship
\begin{equation}
\label{matel}
|<\psi_a|\hat O|\psi_0>|^2=
\hbar\lim_{\Omega\to\Omega_a}(\Omega-\Omega_a)
\overline{<\psi'|\hat O|\psi'>\exp(-i\Omega t)},
\end{equation}
where $\psi_0$ and $\psi_a$ are the stationary wave functions of the
unperturbed ground and excited states; $\psi'$ is the wave function
of the perturbed ground state, $\Omega_a=(E_a-E_0)/\hbar$ are the
normal frequencies, the bar means averaging over a time interval much
larger than $1/\Omega$.

To calculate the {\bf magnetic} transition probability, it is necessary
to excite the system by the following external field:
\begin{equation}
\label{Magnet}
\hat O_{\lambda\mu'}=\mu_N\left(g_s\hat\bS/\hbar-ig_l\frac{2}{\lambda+1}[\br\times\nabla]\right)
\nabla(r^{\lambda}Y_{\lambda\mu'}), \quad 
\mu_N=\frac{e\hbar}{2mc}.
\end{equation}
Here $g_l=1,$ $g_s=5.5856$ for protons and $g_l=0,$ $g_s=-3.8263$ for neutrons.
We are interested in the dipole operator ($\lambda=1,\,\mu'=1$). In cyclic
coordinates it looks like
\begin{equation}
\label{Magnet1}
\hat O_{11}=
\mu_N\sqrt{\frac{3}{4\pi}}\left[g_s\hat S_{1}/\hbar
-g_l\sqrt2\sum_{\nu,\sigma}
C_{1\nu,1\sigma}^{11}r_{\nu}\nabla_{\sigma}\right]
\end{equation}
Its Wigner transformation is
$$(O_{11})_W=
\mu_N\sqrt{\frac{3}{4\pi}}\left[g_s\hat S_{1}
-i\sqrt2g_l\sum_{\nu,\sigma}
C_{1\nu,1\sigma}^{11}r_{\nu}p_{\sigma}\right]/\hbar,$$
 For the matrix element we have
\begin{eqnarray}
\label{psiO}
<\psi'|\hat O_{11}|\psi'>=
\mu_N\sqrt{\frac{3}{4\pi}}\left[
g_s\hat <\psi'|S_{1}|\psi'>
-i\sqrt2L_{11}^{\rm p+}\right]/\hbar
\nonumber\\
=\mu_N\sqrt{\frac{3}{2\pi}}\left[-\frac{\hbar}{2}
(g_s^{\rm n}F^{d\rm n}+g_s^{\rm p}F^{d\rm p})
-iL_{11}^{\rm p+}\right]/\hbar
\nonumber\\
=\mu_N\sqrt{\frac{3}{8\pi}}\left[-\frac{1}{2}
[(g_s^{\rm n}+g_s^{\rm p})F^{d}+(g_s^{\rm n}-g_s^{\rm p})\bar F^{d}]
-\frac{i}{\hbar}(\L_{11}^{+}-\bar \L_{11}^{+})\right]
\nonumber\\
=\mu_N\sqrt{\frac{3}{8\pi}}\left[
\frac{1}{2}(g_s^{\rm p}-g_s^{\rm n})\bar F^{d}
+\frac{i}{\hbar}\bar \L_{11}^{+}
+\frac{i}{\hbar}[g_s^{\rm n}+g_s^{\rm p}-1]\L_{11}^{+}
\right].
\end{eqnarray}
Deriving (\ref{psiO}) we have used equilibrium relations (\ref{equi3})
and $2iL^+_{11}=-\hbar F^{d}$,
which follows from the angular momentum conservation.

Due to the external field one of the dynamical equations of (\ref{iv})
becomes inhomogeneous:
\begin{equation}
\label{Rext}
     \dot {\bar\R}^{+}_{21}
-\frac{2}{m}\bar\L_{21}^{+}
+i\hbar\frac{\eta}{2}[\bar\R_{21}^-
+2\bar\R^{u}_{22}+
2\sqrt{\frac{3}{2}}\bar\R^{d}_{20}]=
i\frac{3\mu_N}{4\hbar\sqrt\pi}R_{20}(eq)\exp^{i\Omega t}.
\end{equation}
Solving the inhomogeneous set of equations (\ref{iv}) in the isovector
and isoscalar cases one can find the required values of
$\bar \L_{11}^{+}\,$, $\L_{11}^{+}$ and $\bar F^{d}$ and calculate
B(M1) factors for all (isovector and isoscalar) excitations.

To calculate the {\bf electric} transition probability, it is 
necessary to excite the system with the external field operator
\begin{equation}
\label{O2mu}
\hat O_{2\mu'}=er^2Y_{2\mu'}=\beta \{r\otimes r\}_{2\mu'},
\end{equation}
where $\beta=e\sqrt\frac{15}{8\pi}$.
Its Wigner transform is identical to (\ref{O2mu}):
$(O_{2\mu'})_W=\beta \{r\otimes r\}_{2\mu'}$.
We consider the case $\mu'=1$. The matrix element is given by
\begin{equation}
\label{psiO2}
<\psi'|\hat O_{21}|\psi'>=\beta R_{21}^{\rm p+}=
\frac{1}{2}\beta (R_{21}^+-\bar R_{21}^+)
=\frac{1}{2}\beta (\R_{21}^+-\bar \R_{21}^+).
\end{equation}
The external field yields two inhomogeneous equations of (\ref{iv}):
\begin{eqnarray}
\label{ivext}
     &&\dot {\bar\L}^{+}_{21}
-\frac{1}{m}\bar\P_{21}^{+} 
+\left[m\,\omega^2
-\frac{2}{\sqrt3}\alpha\chi_0R^{+}_{00}(eq)
+\frac{1}{\sqrt6}(1+\alpha)\chi_0 R^{+}_{20}(eq)\right]\bar\R^{+}_{21}
\nonumber\\
&&\quad\qquad+i\hbar\frac{\eta}{2}[\bar\L_{21}^-
+2\bar\L^{u}_{22}+2\sqrt{\frac{3}{2}}\bar\L^{d}_{20}]
=\frac{\beta}{\sqrt3}\left(\frac{\sqrt2}{4}R_{20}(eq)-R_{00}(eq)\right)
\exp^{i\Omega t},
\nonumber\\
     &&\dot {\bar\L}^{+}_{11}
+\sqrt{\frac{3}{2}}(1-\alpha)\chi_0 R_{20}(eq)\bar\R^{+}_{21}
+i\hbar\frac{\eta}{2}[\bar\L_{11}^- 
+\sqrt2\bar\L^{d}_{10}]=
-\frac{\beta\sqrt6}{4}R_{20}(eq)\exp^{i\Omega t}.
\end{eqnarray}
Solving the inhomogeneous set of equations (\ref{iv}) in the isovector
and isoscalar cases one can find the required values of
$\bar \R_{21}^{+}$ and $\R_{21}^{+}$ and calculate
B(E2) factors for all (isovector and isoscalar) excitations.

\subsection{Isovector excitations}

The energies and excitation probabilities obtained by the solution of
the isovector set of equations (\ref{iv}) are given in
Table 1. These results will be discussed by comparing them with some 
simplified cases, that will allows one to understand the physical 
nature of the considered excitations.

%\newpage
{\bf Tables 1, 2.} Energies and excitation probabilities of the 
isovector system, calculated for $^{164}$Er with two values of the 
spin orbit interaction constant $\eta$. The quantum numbers 
(including indices $+,\,-,\,u,\,d$) of variables responsible for 
the generation of the present level are shown in a first column of
Table 2.
\begin{center}
{\bf Table 1.} \hspace{1mm}$\eta=0.361$ MeV\hspace{5cm}
{\bf Table 2.}\hspace{2mm}$\eta=0$\hspace*{2cm} \\
%{\bf Isovector eigenfrequencies}\\
\begin{tabular}{c|c|c}
%\multicolumn{3}{c}{$\eta=0.361$ MeV}  \B \\ 
\hline        
\multicolumn{1}{l|}{ $E_{\rm iv}$, MeV}  %\T 
 &
\multicolumn{1}{c|}{ $B(M1),\ \mu_N^2$ }     &
\multicolumn{1}{c}{ $B(E2),  \ B_W$}  \B \\ 
\hline\hline%\T  
        ~1.61     & ~3.54   & ~0.12      \\ 
        ~2.18     & ~5.33   & ~1.02      \\
        12.80     & ~0.01   & ~0.04      \\ 
        14.50     & ~0.01   & ~0.03      \\
        16.18     & ~0.02   & ~0.18      \\ 
        16.20     &     0   &    0      \\ 
        20.59     & ~2.78   & 35.45    \B\\
        ~0.26\,i  & -5.43\,i& ~0.0\,i    \\
 \hline
 \end{tabular}
\hspace*{0.5cm}
\begin{tabular}{l||c|c|c}
%\multicolumn{4}{c}{$\eta=0$ MeV}  \B \\ 
\hline          
\multicolumn{1}{l||}{$(\lambda,\mu)^s$} &
\multicolumn{1}{l|}{ $E_{\rm iv}$, MeV}  %\T 
 &
\multicolumn{1}{c|}{ $B(M1),\ \mu_N^2$ }     &
\multicolumn{1}{c}{ $B(E2),\ B_W$} \B \\ 
\hline\hline%\T  
  (1,1)$^-$ &  ~1.70     &    0  &    0       \\ 
  (1,1)$^+$ &  ~2.07     & 9.03  & ~1.17       \\
  (0,0)$^d$ &  12.81     &    0  &    0       \\ 
  (2,1)$^-$ &  14.50     &    0  &    0       \\
  (2,2)$^u$ &  16.20     &    0  &    0       \\ 
  (2,0)$^d$ &  16.20     &    0  &    0       \\ 
  (2,1)$^+$ &  20.55     & 2.81  & 35.70   \B\\
  (1,0)$^d$ &      0     &    -  &    -       \\
 \hline
 \end{tabular}
 \end{center}

\subsubsection{The limit $\eta=0$}

 In the case of vanishing spin orbit potential, $\eta=0$, the set of 
equations (\ref{iv}) splits into four independent subsets. 

{\bf The first one} is
\begin{eqnarray}
\label{iv+}
     \dot {\bar\L}^{+}_{21}&=&
\frac{1}{m}\bar\P_{21}^{+}-
\left[m\,\omega^2
-\frac{2}{\sqrt3}\alpha\chi_0 R_{00}(eq)
+\frac{1}{\sqrt6}(1+\alpha)\chi_0 R_{20}(eq)\right]\bar\R^{+}_{21},
\nonumber\\
     \dot {\bar\L}^{+}_{11}&=&
-\sqrt{\frac{3}{2}}(1-\alpha)\chi_0 R_{20}(eq)\bar\R^{+}_{21},
\nonumber\\
     \dot {\bar\R}^{+}_{21}&=&
\frac{2}{m}\bar\L_{21}^{+},
\nonumber\\
     \dot {\bar\P}^{+}_{21}&=&
-[2m\,\omega^2+\sqrt{\frac{2}{3}}\chi_0 R_{20}(eq)]\bar\L^{+}_{21}
+\sqrt6\chi_0 R_{20}(eq)\bar\L^{+}_{11}.
\end{eqnarray}
This set of equations for spin-scalar variables coincides with
the set of equations (29) of \cite{BaSc} and describes the joint
dynamics of the "standard" nuclear scissors mode and the $\mu=1$ 
branch of IVGQR.
 Remembering that $\chi_0=6\kappa_0,$ 
$R_{2\mu}=Q_{2\mu}/\sqrt6,$ $Q_{20}=Q_{00}\frac{4}{3}\delta$ and 
taking the self consistent value 
$\kappa_0=-\frac{m\bar\omega^2}{4Q_{00}}$ \cite{Ann} we find
the energies of these modes
\begin{equation}
\label{Omeg2}
\Omega^2_{\pm}=\bar{\omega}^2(2-\alpha)(1+\delta/3)
\pm \sqrt{\bar{\omega}^4(2-\alpha)^2(1+\delta/3)^2
-4\bar{\omega}^4(1-\alpha)\delta^2}.
\end{equation}
Usually \cite{BaSc} we choose $\alpha=-2$. Then
\begin{equation}
\label{OmSci}
\Omega^2_{\pm}=4\bar{\omega}^2\left(1+\frac{\delta}{3}
\pm \sqrt{(1+\frac{\delta}{3})^2
-\frac{3}{4}\delta^2}\right).
\end{equation}

{\bf The second subset} reads
\begin{eqnarray}
\label{iv-}
     \dot {\bar\L}^{-}_{21}&=&
\frac{1}{m}\bar\P_{21}^{-}-
[m\,\omega^2+\frac{1}{\sqrt6}\chi_0 R_{20}(eq)]\bar\R^{-}_{21},
\nonumber\\
     \dot {\bar\L}^{-}_{11}&=&
-\sqrt{\frac{3}{2}}\chi_0 R_{20}(eq)\bar\R^{-}_{21},
\nonumber\\
     \dot {\bar\R}^{-}_{21}&=&
\frac{2}{m}\bar\L_{21}^{-},
\nonumber\\
     \dot {\bar\P}^{-}_{21}&=&
-[2m\,\omega^2+\sqrt{\frac{2}{3}}\chi_0 R_{20}(eq)]\bar\L^{-}_{21}
+\sqrt6\chi_0 R_{20}(eq)\bar\L^{-}_{11}.
\end{eqnarray}
In the absence of spin dependent forces one could naively expect that
this set of equations should be identical with (\ref{iv+}). However we
observe an essential difference between the two, the difference being
determined by the variation of the mean field (the variable $Z_{21}^+$).
This fact can easily be understood by comparing the first and second
equations of (\ref{quadr}). The terms responsible for the contribution
of the mean field are $\{Z_2^+\otimes R_2^+\}_{\lambda\mu}$ and
$\{Z_2^+\otimes R_2^-\}_{\lambda\mu}$. Their variations are
$\{\delta Z_2^+\otimes R_2^+(eq)+Z_2^+(eq)\otimes\delta R_2^+\}_{\lambda\mu}$ and
$\{\delta Z_2^+\otimes R_2^-(eq)+Z_2^+(eq)\otimes\delta R_2^-\}_{\lambda\mu}$ 
respectively.
We assumed that $R_{2\mu}^-(eq)=0,$ so the contribution of the mean 
field variation $\delta Z_{2\mu}^+$ in the second equation disappears
and we arrive to the announced difference. In other words, when the
particles with different directions of spins move in phase, the mean 
field variations generated by their motion add and, 
when they move out of phase, the respective mean field variations
annihilate one another.

The set of equations for spin-vector variables (\ref{iv-}) has the
following eigenfrequencies:
\begin{eqnarray}
\Omega^2_{\pm}=2\bar\omega^2
\left(1+\frac{\delta}{3}\pm \sqrt{(1+\frac{\delta}{3})^2-
\delta^2}\,\right).
\label{Omeg2fin}
\end{eqnarray}
The numerical estimates for $^{164}Er$ are shown in Table 2. The low
lying level has $E^-_{11}(low)=\hbar\Omega_-=1.70$ MeV and the high 
lying level has $E^-_{21}(high)=\hbar\Omega_+=14.50$ MeV. So, in 
addition to the well known ("standard") scissors mode with energy 
$E^+_{11}=2.07$ MeV, describing the relative rotational vibrations
of all (spin up + spin down) neutrons with respect of all protons, we
get one more scissors mode, which is generated by the new type of
the nuclear collective motion -- the relative rotational vibrations 
of spin up nucleons with respect of spin down nucleons. The inclusion 
of the spin orbital interaction only slightly shifts these levels 
apart (compare Tables 1 and 2): 1.70 $\to$ 1.61
and 2.07 $\to$ 2.18. However, the main role of the spin orbital 
interaction consists in the excitation of the new scissors mode. 
This
mode is not excited directly by the electromagnetic field (see formula
(\ref{Rext})). It can be excited only indirectly, via the coupling with
the "standard" scissors mode by means of the spin orbital interaction.

The new scissors mode is generated by the variable $\bar\L^-_{11}$ 
responsible for the relative rotational motion. It is easy to see by
removing this variable from the set of equations (\ref{iv-}). In this
case its characteristic equation gives only one level with energy
$E=2\hbar\bar\omega\sqrt{1+\delta/3}$, which practically 
coincides with $E^-_{21}(high)$ and lies in between  the isoscalar and
isovector giant quadrupole resonances. From the physical 
interpretation of variables $\bar\R^-_{21}, \bar\P^-_{21}, 
\bar\L^-_{21},$ responsible for this mode, it follows that it 
describes rather complicate motion: the proton system oscillates out
of phase with the neutron system, whereas inside of each system spin 
up nucleons oscillate out of phase with spin down nucleons. In the 
absence of the proper
residual interaction it has the standard shell model value of the
energy $E=2\hbar\omega(\delta)$. So, the new scissors mode is
accompanied by the high lying excitation, which can be called the 
spin-vector giant quadrupole resonance. As one can see, the situation
is quite similar to that of the "standard" scissors mode which exists 
only together \cite{BaSc} with isovector giant quadrupole resonance.

Let us point to the interesting feature of the new scissors mode. As
one knows, the variable $\bar\L^-_{11}$ is spin-vector (see the text
after eq. (\ref{Varis})), i.e. it describes the relative rotational
oscillations of spin up nucleons with respect of spin down nucleons
(spin scissors). This means that we have proton spin scissors and 
neutron spin scissors. At the same time $\bar\L^-_{11}$ is the 
isovector variable, i.e. it describes the relative rotational 
oscillations of protons with respect of neutrons. So, the resulting
motion described by the variable $\bar\L^-_{11}$ looks rather complex
-- proton spin scissors vibrate in a rotational way (like scissors)
with respect of neutron spin scissors, i.e. we have the scissors made
of two other scissors!

It is necessary also to stress the important role of the Fermi surface
deformation (variable $\bar\P^-_{21}$). Here the situation is exactly the
same as in the case of the "standard" scissors mode. If one removes the
variable $\bar\P^-_{21}$ from the set of equations (\ref{iv-}), one gets
a zero value for $E^-_{11}(low)$ and a considerably changed
expression for the high lying mode - 
$E^-_{21}(high)=\hbar\bar\omega\sqrt{2(1+\delta/3)}$.

{\bf The third subset reads}
\begin{eqnarray}
\label{ivu}
     \dot {\bar\L}^{u}_{22}&=&
\frac{1}{m}\bar\P_{22}^{u}-
[m\,\omega^2
-\sqrt{\frac{2}{3}}\chi_0R_{20}(eq)]\bar\R^{u}_{22},
\nonumber\\
     \dot {\bar\R}^{u}_{22}&=&
\frac{2}{m}\bar\L_{22}^{u},
\nonumber\\
     \dot {\bar\P}^{u}_{22}&=&
-[2m\,\omega^2-2\sqrt{\frac{2}{3}}\chi_0 R_{20}(eq)]\bar\L^{u}_{22}.
\end{eqnarray}
Its solution gives 
$E_{22}=2\hbar\bar\omega\sqrt{(1+\frac{4}{3}\delta)}$.

{\bf The fourth subset reads}
\begin{eqnarray}
\label{ivd}
     \dot {\bar\L}^{d}_{20}&=&
\frac{1}{m}\bar\P_{20}^{d}-
[m\,\omega^2
+\sqrt{\frac{2}{3}}\chi_0 R_{20}(eq)]\bar\R^{d}_{20}
+\frac{2}{\sqrt3}\chi_0 R_{20}(eq)\bar\R^{d}_{00},
\nonumber\\
     \dot {\bar\R}^{d}_{20}&=&
\frac{2}{m}\bar\L_{20}^{d},
\nonumber\\
     \dot {\bar\P}^{d}_{20}&=&
-[2m\,\omega^2+2\sqrt{\frac{2}{3}}\chi_0 R_{20}(eq)]\bar\L^{d}_{20}
+\frac{4}{\sqrt3}\chi_0 R_{20}(eq)\bar\L^{d}_{00},
\nonumber\\
     \dot {\bar\L}^{d}_{00}&=&
\frac{1}{m}\bar\P_{00}^{d}-m\,\omega^2\bar\R^{d}_{00}
+\frac{2}{\sqrt3}\chi_0 R_{20}(eq)\bar\R^{d}_{20},
\nonumber\\
     \dot {\bar\R}^{d}_{00}&=&
\frac{2}{m}\bar\L_{00}^{d},
\nonumber\\
     \dot {\bar\P}^{d}_{00}&=&
-2m\,\omega^2\bar\L^{d}_{00}
+\frac{4}{\sqrt3}\chi_0 R_{20}(eq)\bar\L^{d}_{20}.
\end{eqnarray}
Its solution gives 
$E_{20}=2\hbar\bar\omega\sqrt{(1+\frac{4}{3}\delta)}$ and
$E_{00}=2\hbar\bar\omega\sqrt{(1-\frac{2}{3}\delta)}$.

According to the physical interpretation of variables of the third and
fourth subsets
they describe the spin-flip modes: giant quadrupole and
monopole spin-flip resonances with energies $E_{20}=E_{22}=16.20$ MeV
and $E_{00}=12.81$ MeV respectively (Table 2). Our Hamiltonian does
not contain any forces, which can change the direction of the spin, so
these modes have zero values of excitation probabilities and the 
standard shell model values of energies $E=2\hbar\omega(\delta)$. The 
spin orbital interaction changes this result only negligibly.

And finally, a few words about the solution with the imaginary 
eigenfrequency (last line of Table 1). From where does it appear? We 
carefully checked that it is not a mistake of calculation or 
computation, because the energy weighted sum rule (see Appendix C) is
fulfilled, the contribution of this unexpected mode being $\sim$1.5\%. 
What could be its physical interpretation? The time dependence of 
variables in
this mode is $e^{-\Omega_0t}$ ($\Omega_0=0.26$ MeV), i.e. this
excitation is unstable, i.e. decaying. 
%This means that at the frequency of the 
%external (perturbing) field $\Omega=\Omega_0$ the system is able to
%"swallow"  some portion of the energy coming from outside,
%but this excitation begins to decay immediately because of the absence
%of the proper restoring force. 
This may be connected with
the neglect by higher order moments. One can hope also that the 
inclusion of the spin-spin or spin-multipole residual interaction 
will produce the necessary restoring force and stabilize the mode.

\subsubsection{The approximation 
$f^{\downarrow\uparrow}=f^{\uparrow\downarrow}=0$}

The set of equations for the variables $X^{\pm}$ is obtained from 
(\ref{iv}) by setting variables $X^u, X^d$ to zero. Its solution gives
energies and excitation probabilities given in the Table 3. The 
comparison with the exact solution shows that the approximation of two
(spin up and spin down) "liquids" can be used with great care - it 
works very well to describe energies and $B(E2)$ factors, but leads to
erroneous results for $B(M1)$ factors because the essential contribution
into their values stems from the (now neglected) variable $\bar F^d$.

{\bf Table 3.} Energies and excitation probabilities of the isovector
system, calculated for $^{164}$Er with $\eta=0.361$ MeV in the approximation 
$f^{\downarrow\uparrow}=f^{\uparrow\downarrow}=0$.
\begin{center}
\begin{tabular}{c|c|c}
%\multicolumn{3}{c}{$\eta=0.361$ MeV}  \B \\ 
\hline        
\multicolumn{1}{l|}{ $E_{\rm iv}$, MeV}  %\T 
 &
\multicolumn{1}{c|}{ $B(M1),\ \mu_N^2$ }     &
\multicolumn{1}{c}{ $B(E2),\ B_W$}  \B \\ 
\hline\hline%\T  
        ~1.62     & ~1.30  & ~0.16      \\ 
        ~2.15     & ~7.74  & ~1.01      \\
        14.50     &     0  & ~0.03      \\
        20.59     & ~2.81  & 35.66    \B\\
 \hline
 \end{tabular}
 \end{center}

\subsection{Isoscalar case}

The energies and excitation probabilities obtained by the solution of
the isoscalar set of equations (\ref{iv}) are demonstrated in the 
Table 4. Analogously to the isovector case the results will be discussed
in comparison with $\eta=0$ limit, what helps to understand their physical 
nature.

{\bf Tables 4, 5.} Energies and excitation probabilities of the 
isoscalar system, calculated for $^{164}$Er with two values of the 
spin orbital interaction constant $\eta$. The quantum numbers 
(including indices $+,\,-,\,u,\,d$) of variables responsible for 
the generation of the present level are shown in a first column of
Table 5.
\begin{center}
{\bf Table 4.} \hspace{1mm}$\eta=0.361$ MeV\hspace{5cm}
{\bf Table 5.}\hspace{2mm}$\eta=0$\hspace*{2cm} \\
\begin{tabular}{c|c|c}
%\multicolumn{3}{c}{$\eta=0.361$ MeV}  \B \\ 
\hline        
\multicolumn{1}{l|}{ $E_{\rm is}$, MeV}  %\T 
 &
\multicolumn{1}{c|}{ $B(M1),\ \mu_N^2$ }     &
\multicolumn{1}{c}{ $B(E2),\ B_W$}  \B \\ 
\hline\hline%\T  
        ~1.73     & -0.07     & ~~1.12       \\
        ~0.39     & ~0.24     & 117.19      \\ 
        12.83     &     0     & ~~0.66      \\
        14.51     &     0     & ~~0.12      \\ 
        16.20     &     0     &     0      \\ 
        16.22     &     0     & ~~0.20    \B\\
        10.28     &     0     & ~66.50      \\ 
        ~0.20\,i  & -0.12\,i  & ~30.30\,i      \\
 \hline
 \end{tabular}
\hspace*{0.5cm}
\begin{tabular}{l||c|c|c}
%\multicolumn{4}{c}{$\eta=0$ MeV}  \B \\ 
\hline          
\multicolumn{1}{l||}{$(\lambda,\mu)^s$} &
\multicolumn{1}{l|}{ $E_{\rm is}$, MeV}  %\T 
 &
\multicolumn{1}{c|}{ $B(M1),\ \mu_N^2$ }     &
\multicolumn{1}{c}{ $B(E2),\ B_W$} \B \\ 
\hline\hline%\T  
  (1,1)$^-$ &  ~1.70     & 0  & 0       \\
  (1,1)$^+$ &    0       & -  & -       \\ 
  (0,0)$^d$ &  12.81     & 0  & 0       \\
  (2,1)$^-$ &  14.51     & 0  & 0       \\ 
  (2,2)$^u$ &  16.20     & 0  & 0       \\ 
  (2,0)$^d$ &  16.20     & 0  & 0 \B\\
  (2,1)$^+$ &  10.33     & 0  & 67.47       \\ 
  (1,0)$^d$ &    0       & -  & -       \\
 \hline
 \end{tabular}
 \end{center}

\subsubsection{The limit of vanishing spin-orbit potential ($\eta=0$)}

In the isoscalar case the set of equations (\ref{iv+}) is transformed
into
\begin{eqnarray}
\label{is+}
     \dot {\L}^{+}_{21}&=&
\frac{1}{m}\P_{21}^{+}-
\left[m\,\omega^2
-\frac{2}{\sqrt3}\chi_0 R_{00}(eq)
+\frac{2}{\sqrt6}\chi_0 R_{20}(eq)\right]\R^{+}_{21},
\nonumber\\
     \dot {\L}^{+}_{11}&=&0,
\nonumber\\
     \dot {\R}^{+}_{21}&=&
\frac{2}{m}\L_{21}^{+},
\nonumber\\
     \dot {\P}^{+}_{21}&=&
-[2m\,\omega^2+\sqrt{\frac{2}{3}}\chi_0 R_{20}(eq)]\L^{+}_{21}
+\sqrt6\chi_0 R_{20}(eq)\L^{+}_{11}.
\end{eqnarray}
These equations coincide with the set of equations (24) of \cite{BaSc}.
They describe the joint dynamics of the $\mu=1$ branch of ISGQR and 
the orbital angular momentum (its $l_1$ projection). In the absence
of spin the orbital angular momentum is conserved 
($\frac{d}{dt}<\hat l_1>=-i\sqrt2\dot{\L}^{+}_{11}=0$) as it 
should be. However, in the case with spin the situation is changed 
radically. Now the orbital angular momentum and spin can vibrate out
of phase keeping their sum (the total angular momentum) unchanged.
This new circumstance reveals itself in the appearance of the new low
lying mode which describes the relative motion of the orbital angular
momentum and spin of the nucleus (see Table 4). According to
our calculations this mode (E=0.39 MeV) has essentially electric
character with $B(E2)=117~W.u.$ demonstrating also a small
admixture of the magnetic properties with $B(M1)=0.24 \mu_N^2$.

As it is seen, the set of equations (\ref{iv-}) does not depend on 
$\alpha$, so it is not changed in the isoscalar case. Naturally it
has just the same eigenvalues: $E(low)=1.70$ MeV, $E(high)=14.50$
MeV. The collective motion corresponding to these two modes is more 
simple, than in the isovector case. The variable $\L^-_{11}$, 
responsible for the low lying excitation, describes the real spin 
scissors: all spin up nucleons (protons together with neutrons)
oscillate rotationally out of phase with all spin down 
nucleons. 
This mode has B(M1)=0 because magnetic moments produced by 
two parts of spin scissors annihilate one another. The variables
$\R^-_{21}, \P^-_{21}, \L^-_{21},$ responsible for the high lying
excitation, describe out of phase oscillations of all spin up nucleons
(protons together with neutrons) with respect of all spin down 
nucleons. 
The inclusion of the spin orbital interaction influences 
practically only on the low-lying mode -- its energy becomes a little
bit higher (1.70 MeV $\to$ 1.73 MeV) and it acquires the nonzero B(E2)
value (B(E2)=1.12 W.u.) demonstrating its electric character. The 
negligibly small negative B(M1) value and the imaginary solution 
(last line of the Table 4) appear, probably, due to the same reasons,
as was already discussed at the end of section 5.1.1.

\subsubsection{The approximation 
$f^{\downarrow\uparrow}=f^{\uparrow\downarrow}=0$}

Removing the variables $X^u, X^d$ from (\ref{iv}) with $\alpha=1$ we
obtain the isoscalar set of equations for $X^{\pm}$ variables. The 
results of calculations are shown in the Table 6. It is readily seen,
that approximate results reproduce rather well the exact ones 
(Table 4) except for one case: instead of the excitation with the energy
E=0.39 MeV, which is generated by the relative motion of the orbital
angular momentum and the spin, we obtain an imaginary solution. This
is not surprising, since after removing the variable $F^d$, one
can not ensure the conservation of the total angular momentum (see 
section 4.3).

{\bf Table 6.} Energies and excitation probabilities of the isoscalar
system, calculated for $^{164}$Er with $\eta=0.361$ MeV in the 
approximation $f^{\downarrow\uparrow}=f^{\uparrow\downarrow}=0$.
\begin{center}
\begin{tabular}{c|c|c}
%\multicolumn{3}{c}{$\eta=0.361$ MeV}  \B \\ 
\hline        
\multicolumn{1}{l|}{ $E_{\rm iv}$, MeV}  %\T 
 &
\multicolumn{1}{c|}{ $B(M1),\ \mu_N^2$ }     &
\multicolumn{1}{c}{ $B(E2),\ B_W$}  \B \\ 
\hline\hline%\T  
  0.18\,i   & 0.79\,i & -212.4\,i       \\ 
  ~1.72     & 0.08    & ~0.99       \\
  10.32     &    0    & 67.34       \\
  14.51     &    0    & ~0.12   \B\\ \hline 
 \end{tabular}
 \end{center}

\section{Conclusion}

In this work, the WFM method is applied for a first time 
to solve the TDHF equation 
including spin dynamics. The model Hamiltonian consists of 
a harmonic oscillator with
spin orbit mean field potential plus quadrupole-quadrupole residual 
interaction. Spin dependent
collective variables are defined and a corresponding closed set of nonlinear
dynamical equations is set up. The equations are solved in  
small amplitude approximation. Two isovector and two isoscalar low
lying eigenfrequencies and five isovector and five isoscalar high 
lying eigenfrequencies have been found. Three low lying levels correspond
to the excitation of a new kind of mode, unknown earlier. 
For example the isovector level with energy E=1.61 MeV describes the
rotational oscillations of nucleons with the spin projection "up"
with respect of nucleons with the spin projection "down", i.e. one
can talk of a nuclear spin scissors mode. Thus the experimentally 
observed group of 1$^+$
peaks in the interval 2-4 MeV, associated usually with the nuclear 
scissors mode, in reality consists of the excitations of 
the  "spin" scissors mode together with the "standard" scissors mode.
It is necessary to point out an interesting feature of
this new mode. Being of isovector nature, it describes also the motion of 
protons with respect to neutrons and, therefore, the resulting motion
can be characterized as the rotational oscillations of proton spin
scissors with respect of neutron spin scissors. In the analogous 
isoscalar mode  with the energy E=1.73 MeV the proton and neutron 
spin scissors move in phase. This mode has B(M1)=0 because magnetic
moments produced by two different parts of proton spin scissors
annihilate each other. A small B(E2) factor  is explained by the 
coupling with the ISGQR.

One more new low lying mode with energy E=0.39 MeV  is generated
by the out of phase oscillations of the orbital angular momentum and 
spin of the nucleus. The rather big B(E2) factor of this mode still waits for 
its explanation.

There are ten high lying excitations, the two of them being really new:
isovector and isoscalar "spin-vector" resonances with 
energies E=14.5 MeV. Further six high lying modes can be interpreted as 
spin-flip ones.
In the absence of residual interactions all these eight
modes have very small excitation probabilities and standard
shell model energies E=2$\hbar\omega(\delta)$. 

In the light of the above results, the study of all discussed (low 
and high lying) excitations with proper residual interactions
included will be the natural continuation of this work. It is known
very well that pairing is very important for the correct description
of the "standard" scissors. Therefore one has to take into 
account pair correlations too, i.e. to move from the TDHF equations to 
TDHFB equations. This also will be a task for the future.

\section*{Acknowledgments}

The useful discussions with M. Urban are gratefully acknowledged.

\section{Appendix A}

According to the definition \cite{Ring} of Wigner transformation one
finds
\begin{eqnarray}
h_{l_0}(\br,\bp)=\int\!d^3s e^{-i\bp\bs/\hbar}h_{l_0}(\br+\frac{\bs}{2},
\br-\frac{\bs}{2})
\nonumber\\
=-\frac{\hbar}{2}\eta(\br)(xp_y-yp_x)-\frac{i\hbar}{2}
(x\frac{\partial}{\partial y}-y\frac{\partial}{\partial x})\frac{\hbar}{2}\eta(\br).
\label{WigH}
\end{eqnarray}
Usually we will take $\eta(\br)=\eta(r)$. Then $\frac{\partial}
{\partial x_i}\eta(\br)=\frac{x_i}{r}\frac{\partial\eta}{\partial r}$.
As a result 
$(x\frac{\partial}{\partial y}-y\frac{\partial}{\partial x})\eta(\br)=0$
and
\begin{eqnarray}
h_{l_0}(\br,\bp)=-\frac{\hbar}{2}\eta(r)(xp_y-yp_x).
\label{WiH0}
\end{eqnarray}

In a similar way one finds
\begin{eqnarray}
h_{l_1}(\br,\bp)=\frac{\hbar}{2}\eta(r)[i(xp_z-zp_x)-(yp_z-zp_y)],
\nonumber\\ 
h_{l_{-1}}(\br,\bp)=-\frac{\hbar}{2}\eta(r)[i(xp_z-zp_x)+(yp_z-zp_y)].
\label{WiH1}
\end{eqnarray}

The Wigner transformation of a product of two operators is given \cite{Ring} by
the following formula 
\begin{eqnarray}
(\hat h\hat\rho)_W&=&
h(\br,\bp)\exp(\frac{i\hbar}{2}\stackrel{\leftrightarrow}
{\Lambda})f(\br,\bp)
\nonumber\\
 &=& h(\br,\bp)f(\br,\bp)
+\frac{i\hbar}{2}\{h,f\}-\frac{\hbar^2}{8}
\{\{h,f\}\}+o(\hbar^3),
\label{WigK}
\end{eqnarray}
where 
$\stackrel{\leftrightarrow}{\Lambda}=
\stackrel{\gets}\nabla_r\stackrel{\to}\nabla_p
-\stackrel{\gets}\nabla_p\stackrel{\to}\nabla_r, \quad
\{h,f\}=h\stackrel{\leftrightarrow}{\Lambda}f$ is the Poisson
bracket of functions $h(\br,\bp)$ and $f(\br,\bp),\quad$
$\{\{h,f\}\}=
h(\br,\bp)({\stackrel{\leftrightarrow}{\Lambda}})^2f(\br,\bp)$
is their double Poisson bracket. 

\section{Appendix B}

{\bf Integrals of motion}.
\begin{eqnarray}
{\bar\L}^{+}_{11}&=&-i\,\frac{2(1-\alpha)\chi_0}{\hbar\eta R^{+}_{20}(eq)}
\sqrt{\frac{3}{2}}C^d
+i\,\frac{\hbar}{2}\,\bar F^d,\nonumber \\
{\bar\P}^d_{20}&=&m\left(m\omega^2-\sqrt{\frac{2}{3}}\chi_0 R^{+}_{20}(eq)\right)
\left(\sqrt{\frac{3}{2}}\bar R^u_{22}-\sqrt{2}\bar R^d_{00}
-\bar R^d_{20}\right)+\sqrt{\frac{3}{2}} \bar P^u_{22}-\sqrt{2}\bar P^d_{00},\nonumber \\
{\bar\L}^{+}_{21}&=&i\,\frac{2}{\hbar\eta}\left[\frac{1}{m}\bar P^u_{22}
+\left(m\omega^2-\sqrt{\frac{2}{3}}\chi_0 R^{+}_{20}(eq)\right)
\left(\bar R^u_{22}-C^d\right)\right]\nonumber \\
&-&i\,\frac{2}{\hbar\eta}\left(m\omega^2-\frac{2\alpha \chi_0}{\sqrt{3}} R^{+}_{00}(eq)
+\frac{(1+\alpha)\chi_0}{\sqrt{6}} R^{+}_{20}(eq)\right)\,C^d\nonumber \\
&-&i\,\frac{\hbar\,\eta\, m}{4}\left[\left(\bar R^{-}_{21}+2\bar R^{u}_{22}-3C^d)\right)
+\frac{3\sqrt{2}}{m\chi_0 R^{+}_{20}(eq)}\left(\bar P^d_{00}+m\omega^2 \bar R^d_{00}\right)
\right],\quad\mbox{where}\nonumber \\ 
C^d&\equiv&\sqrt{\frac{2}{3}}\bar R^d_{20}
- \frac{\left(\bar P^d_{00}+m\omega^2 \bar R^d_{00}\right)}{\sqrt{2}m\,\chi_0\,R^{+}_{20}(eq)}.
\nonumber
\end{eqnarray}
 
\section{Appendix C}

{\bf The sum rule for magnetic transitions.}

Magnetic operator 
\begin{eqnarray}
\hat O_{1\phi}=\sqrt{\frac{3}{4\pi}}(g_l\hat l_{\phi} 
+g_s\hat S_{\phi})\mu_N/\hbar
\label{Magn}
\end{eqnarray}
obeys the following sum rule
\begin{eqnarray}
\label{sumrule}
\sum_{\nu}(E_{\nu}-E_0)
(|<\nu|\hat O_{11}|0>|^2+|<\nu|\hat O_{1-1}|0>|^2)
=
-<0|[\hat O_{11},[H,\hat O_{1-1}]]|0>.
\end{eqnarray}
The oscillator part $h_{osc}$ of the Hamiltonian (\ref{Ham}) commutes
with $\hat S_{\phi}$, so we can use for the double commutator of
$\hat O_{1\phi}$ with $h_{osc}$ the result found in \cite{BaSc}:
\begin{equation}
\label{commut}
[\hat O_{1\phi},[h_{osc},\hat O_{1\phi'}]]=
\frac{15}{2\pi}\bar\chi\sum_i^Z
\sum_j^N\sum_{\nu,\sigma,\epsilon}(-1)^{\nu}C_{2\nu,2\sigma}^{1\phi}
C_{2-\nu,2\epsilon}^{1\phi'}\{r_i\otimes r_i\}_{2\epsilon}
\{r_j\otimes r_j\}_{2\sigma}\mu_N^2.
\end{equation}
Its ground state matrix element is \cite{BaSc}
\begin{equation}
\label{Rfin}
<0|[\hat O_{11},[h_{osc},\hat O_{1-1}]]|0>=
\frac{9}{16\pi}\bar\chi (R_{2 0}^{eq})^2\mu_N^2
=-\frac{1-\alpha}{4\pi}Q_{00}m\bar\omega^2\delta^2\mu_N^2.
\end{equation}
Now we have to calculate the double commutator of $\hat O_{1\phi}$ 
with the spin orbital part $h_{ls}$ of the Hamiltonian (\ref{Ham}). 
One has
$$\hat h_{ls}=-\eta(\br)\sum_{\mu=-1}^1(-1)^{\mu}\hat l_{\mu}\hat S_{-\mu},
\quad [\hat l_{\mu},\hat l_{\phi}]= -\sqrt2C^{1\nu}_{1\mu,1\phi}\hat l_{\nu},
\quad [\hat S_{\mu},\hat S_{\phi}]= -\sqrt2C^{1\nu}_{1\mu,1\phi}\hat S_{\nu},$$

$$ [h_{ls},\hat l_{\phi}]=-\eta(\br)\sum_{\mu=-1}^1(-1)^{\mu}
[\hat l_{\mu},\hat l_{\phi}]\hat S_{-\mu}=\sqrt2\eta(\br)\sum_{\mu=-1}
^1(-)^{\mu}C^{1\nu}_{1\mu,1\phi}\hat l_{\nu}\hat S_{-\mu},$$

$$ [h_{ls},\hat S_{\phi}]=-\eta(\br)\sum_{\mu=-1}^1(-1)^{\mu}
\hat l_{\mu}[\hat S_{-\mu},\hat S_{\phi}]=\sqrt2\eta(\br)\sum_{\mu=-1}
^1(-)^{\mu}C^{1\nu}_{1-\mu,1\phi}\hat l_{\mu}\hat S_{\nu},$$

$$ [h_{ls},\hat O_{1\phi}]=\frac{\mu_N}{\hbar}\sqrt{\frac{3}{2\pi}}\eta(\br)\sum_{\mu=-1}
^1(-1)^{\mu}C^{1\nu}_{1\mu,1\phi}(g_l\hat l_{\nu}\hat S_{-\mu}
+g_s\hat l_{-\mu}\hat S_{\nu}),$$

\begin{eqnarray}
\label{sumru1}
[\hat l_{\phi},[h_{ls},\hat O_{1\phi'}]]
&=&-\frac{\mu_N}{\hbar}\sqrt{\frac{3}{\pi}}\eta(\br)\sum_{\mu=-1}^1(-1)^{\mu}C^{1\nu'}_{1\mu,1\phi'}
(g_lC^{1\nu}_{1\phi,1\nu'}\hat l_{\nu}\hat S_{-\mu}
+g_sC^{1\nu}_{1\phi,1-\mu}\hat l_{\nu}\hat S_{\nu'})
\nonumber\\
&=&-3\frac{\mu_N}{\hbar}\sqrt{\frac{3}{\pi}}\eta(\br)(g_l-g_s)\sum_{j,m}\{_{j\,1\,1}^{1\,1\,1}\}
C^{jm}_{1\phi',1\phi}\{\hat l\otimes\hat S\}_{jm},
\end{eqnarray}

\begin{eqnarray}
\label{sumru2}
[\hat S_{\phi},[h_{ls},\hat O_{1\phi'}]]&=&
-\frac{\mu_N}{\hbar}\sqrt{\frac{3}{\pi}}\eta(\br)\sum_{\mu=-1}
^1(-1)^{\mu}C^{1\nu'}_{1\mu,1\phi'}(g_lC^{1\nu}_{1\phi,1-\mu}\hat l_{\nu'}\hat S_{\nu}
+g_sC^{1\nu}_{1\phi,1\nu'}\hat l_{-\mu}\hat S_{\nu}),
\nonumber\\
&=&3\frac{\mu_N}{\hbar}\sqrt{\frac{3}{\pi}}\eta(\br)(g_l-g_s)\sum_{j,m}\{_{j\,1\,1}^{1\,1\,1}\}
(-1)^jC^{jm}_{1\phi',1\phi}\{\hat l\otimes\hat S\}_{jm},
\end{eqnarray}

\begin{eqnarray}
\label{sumru3}
[\hat O_{\phi},[h_{ls},\hat O_{1\phi'}]]
=-\frac{\mu_N^2}{\hbar^2}\frac{9}{2\pi}\eta(\br)\sum_{i=1}^A(g_l^i-g_s^i)
\sum_{j,m}[g_l^i-g_s^i(-1)^j]
\{_{j\,1\,1}^{1\,1\,1}\}C^{jm}_{1\phi',1\phi}\{\hat l^i\otimes\hat S^i\}_{jm}.
\end{eqnarray}

We need the ground state matrix element of this double commutator for $\phi=1,\, \phi'=-1$:
\begin{eqnarray}
\label{Sumls}
<0|[\hat O_{11},[h_{ls},\hat O_{1-1}]]|0>
&=&-\frac{\mu_N^2}{\hbar^2}\frac{9}{2\pi}\sum_{i=1}^A(g_l^i-g_s^i)\sum_{j}[g_l^i-g_s^i(-1)^j]
\{_{j\,1\,1}^{1\,1\,1}\}
\nonumber\\
&&C^{j0}_{1-1,11}<0|\eta(\br)\{\hat l^i\otimes\hat S^i\}_{j0}|0>.
\end{eqnarray}
Averaging of spin variables implies the trace over spin indices, 
which is obviously equal to zero in our case. Therefore the spin 
orbital interaction does not contribute into the sum rule.

{\bf The sum rule for electric transitions} coincides with that of the
paper \cite{BaSc}:
\begin{eqnarray}
\label{sumruleO2}
\sum_{\nu}(E_{\nu}-E_0)
(|<\nu|\hat O_{21}|0>|^2+|<\nu|\hat O_{2-1}|0>|^2)
&=&
-<0|[\hat O_{21},[H,\hat O_{2-1}]]|0>
\nonumber\\
&=&e^2\frac{\hbar^2}{m}\frac{5}{4\pi}
Q_{00}(1+\delta/3).
\end{eqnarray}

\end{document}